# Bayesian hierarchical non-stationary hybrid modeling for threshold estimation in peak over threshold approach

**Quansheng Yue, PH.D. Student**

[a] School of Transportation, Southeast University, Nanjing, China, 211189
[b] Jiangsu Key Laboratory of Urban ITS, Jiangsu Collaborative Innovation Center of Modern Urban Traffic Technologies
Email: yue_qs@seu.edu.cn

**Yanyong Guo, Professor**

[a] School of Transportation, Southeast University, Nanjing, China, 211189
[b] Jiangsu Key Laboratory of Urban ITS, Jiangsu Collaborative Innovation Center of Modern Urban Traffic Technologies
Email: guoyanyong@seu.edu.cn

**Tarek Sayed, Professor**

[c] Department of Civil Engineering
University of British Columbia, Vancouver, BC, Canada V6T 1Z4
Email: tsayed@ubc.ca

**Lai Zheng, Professor**

[d] School of Transportation Science and Engineering
Harbin Institute of Technology, Harbin, China, 150090
Email: zhenglai@hit.edu.cn

**Hao Lyu, PH.D. Student**

[a] School of Transportation, Southeast University, Nanjing, China, 211189
[b] Jiangsu Key Laboratory of Urban ITS, Jiangsu Collaborative Innovation Center of Modern Urban Traffic Technologies
Email: lyu_hao@seu.edu.cn

**Pan Liu, Professor**

[a] School of Transportation, Southeast University, Nanjing, China, 211189
[b] Jiangsu Key Laboratory of Urban ITS, Jiangsu Collaborative Innovation Center of Modern Urban Traffic Technologies
Email: liupan@seu.edu.cn

**Abstract**

Extreme value theory (EVT) has been utilized to estimate crash risk from traffic conflicts with the peak over threshold approach. However, it's challenging to determine a suitable threshold to distinguish extreme conflicts in an objective way. The subjective and arbitrary selection of the threshold in the peak over threshold approach can result in biased estimation outcomes. This study proposes a Bayesian hierarchical hybrid modeling (BHHM) framework for the threshold estimation in the peak over threshold approach. Specifically, BHHM is based on a piecewise function to model the general conflicts with specific distribution while model the extreme conflicts with generalized Pareto distribution (GPD). The threshold for distinguishing general conflicts and extreme conflicts is taken as a model parameter in BHHM which needs to be estimated. The Bayesian hierarchical structure is used to combine traffic conflicts from different sites, incorporating covariates and site-specific unobserved heterogeneity. Five non-stationary BHHM models, including Normal-GPD, Cauchy-GPD, Logistic-GPD, Gamma-GPD, and Lognormal-GPD models, were developed and compared. Traditional graphical diagnostic and quantile regression approaches were also used for comparison. Traffic conflicts collected from three signalized intersections in the city of Surrey, British Columbia were used for the study. The Bayesian approach is employed to estimate the threshold and other parameters in the non-stationary BHHM models. The results show that the proposed BHHM approach could estimate the threshold parameter objectively. The non-stationary BHHM models are able to capture the characteristics of threshold variation across cycles due to the change in traffic status. The Lognormal-GPD model is superior to the other four BHHM models in terms of crash estimation accuracy and model fit. The crash estimates using the threshold determined by the BHHM outperform those estimated based on the graphical diagnostic and quantile regression approaches, indicating the superiority of the proposed threshold determination approach. The findings of this study contribute to enhancing the existing EVT methods for providing a threshold determination approach as well as producing reliable crash estimations.

# 1. Introduction

In recent decades, traffic safety evaluation has heavily relied on traffic crash records to assess safety and formulate corresponding measures (Lord et al., 2005; Washington et al., 2020; Guo et al., 2019a). Despite the advancement of methodologies developed to analyze crash data (Lord and Mannering, 2010; Mannering and Bhat, 2014; Mannering et al., 2016), it cannot be overlooked that the quality of crash data and the limitations associated with crash-based safety analysis methods remain significant concerns (Zheng et al., 2021). To address this issue, alternative safety measures such as traffic conflicts have been proposed and gained considerable popularity for road safety analysis. Compared to crash data, traffic conflicts occur more frequently, are more cost-effective, and are readily available, offering significant advantages (Tarko et al., 2021; Guo et al., 2019b; Arun et al.,2022b; Guo et al., 2024; Alozi and Hussein, 2022; Nazir et al., 2024). Most importantly, as a proactive safety measure, employing traffic conflict analysis allows for preemptive measures to prevent crashes before they happen. To figure out the relationship between traffic conflicts and crashes, traffic safety researchers have developed various approaches, including causal models, statistical models, and extreme value theory (EVT) models (Tarko,2018; Sayed and Zein, 1999; Zheng et al.,2019a; Guo et al., 2019b; Fu et al.,2020; Ali et al., 2022; Ankunda et al., 2024; Hussain et al., 2022). Among these models, EVT models have recently garnered considerable research attention. The use of EVT models is to gain inference about the tails of probability distributions using asymptotically justified models (Behrens et al.,2004). Inference on the tails permits extrapolation beyond the range of the observed data, forming a basis for predictions about the levels of future extremes. As such, it enables extrapolation from observable level to unobservable level, providing a method to estimate crash risk by inferring from frequent conflicts (observable level) to infrequent crashes (unobservable level) (Songchitruksa and Tarko, 2006).

EVT models encompasses two primary methods: peak over threshold which aligns with the generalized Pareto distribution (GPD) (Guo et al., 2020a), and block maximum which aligns with the generalized extreme value distribution (Zheng et al., 2019b). Several studies have indicated that the peak over threshold method is superior to the block maxima method in terms of estimated accuracy and reliability (Zheng et al., 2014a; Borsos et al., 2020). This is because the peak over threshold utilizes more traffic conflict data while the block maxima method tends to discard a significant amount of conflict data. A critical issue with the peak over threshold method is the determination of the threshold to distinguish the extreme conflicts and general conflicts. If the threshold is set too low, it violates the asymptotic basis of the model, resulting in bias estimation. Conversely, if the threshold is set too high, it produces too few excesses, leading to high variance (Coles et al., 2001). Graphical diagnostics such as mean residual life plots and threshold stability plots were the most commonly used methods for threshold determination in the peak over threshold method (Davison and Smith, 1990; Coles et al., 2001). The threshold can be determined when the threshold stability plot is approximately constant and the mean residual life plot is approximately linear. However, it is difficult to judge the linearity and stability accurately. Rules of thumb such as quantile plots method is the other way for threshold determination (DuMouchel, 1983). According to quantile plots method, a specific upper percentile of the data is chosen as the threshold. This percentile is uncertain and is often determined based on expert judgment (DuMouchel, 1983; Roth et al., 2014; Zheng et al., 2019a). For example, DuMouchel (1983) advocates the upper 10% for threshold, while Zheng et al. (2019a) make upper 12% as threshold. Overall, in the process of threshold determination in peak over threshold method, both the graphical diagnostics and rules of thumb are somewhat subjective and arbitrary, potentially leading to significant variations in the estimation results. To address the threshold issue, our previous study proposed a joint gamma and generalized Pareto distribution modeling, where a shifted reciprocal mapping was used to process the traffic conflict data (Yue et al., 2024). However, the shift value was selected by traversal method, remaining a challenging task.

Several previous studies have confirmed that developing non-stationary EVT models to account for covariates that may influence the stationary of extremes leading to better model estimates and improved inference (Zheng et al., 2019b; Fu et al., 2020; Nazir et al.,2023). However, those studies on non-stationary peak over threshold models suffer from the issue of threshold determination as discussed above. Therefore, this study aims to bridge the research gap by developing a Bayesian hierarchical hybrid (BHHM) modeling approach of the non-stationary traffic conflicts for the threshold estimation in peak over threshold approach. The main contributions of this study are as follows: a) it develops a BHHM framework to estimate the threshold in the peak over threshold approach, enabling the objective and quantitative distinction between general conflicts and extreme conflicts; b) it develops five non-stationary BHHM models and compares their performance of threshold estimation in terms of goodness of fit and crash estimation accuracy; and c) it compares the superiority of threshold estimated from the non-stationary BHHM models with those calculated from the graphical diagnostics approach and quantile regression approach in terms of crash estimation.

## 2. Previous Work
### 2.1. Studies on stationary and non-stationary EVT models

The groundbreaking research by Campbell et al. (1996) initiated the use of the EVT approach in road safety analysis. In recent years, EVT has gained widespread adoption in traffic safety research (Songchitruksa and Tarko, 2006; Hussain et al., 2024; Zheng and Sayed, 2019a; Zheng et al., 2019b; Fu et al., 2020; Guo et al., 2020a; Zheng and Sayed, 2020; Fu and Sayed, 2023; Chauhan et al., 2023; Hu et al., 2024; Alozi and Hussein, 2022). The detailed EVT modeling techniques developed and expanded upon by Songchitruksa and Tarko (2006) have significantly propelled the application and advancement of EVT in road safety analysis. In general, the EVT models can be classified into two categories: stationary and non-stationary models. In stationary models, extreme conflicts were assumed to be stationary. Specifically, it is assumed that observed extreme conflicts were from an identical distribution. Guo et al. (2020b) employed the stationary EVT model to evaluate the safety of left-turn bay extension and conducted a comparison with collision-based methods. Borsos et al. (2020) and Zheng et al. (2014a) applied both block maxima and peak over threshold methods to estimate crashes, demonstrating superior performance of the peak over threshold approach over block maxima approach in terms of data utilization, estimation accuracy, and reliability. Guo et al. (2021) integrated microsimulation with the EVT approach for safety evaluation. Several studies developed stationary bivariate EVT models for crash estimation incorporating two conflict indicators (Zheng et al., 2019a; Arun et al., 2021).

Non-stationary models could account for the non-stationarity of extreme conflicts by incorporating factors influencing crash occurrences. Zheng et al. (2019b) employed a Bayesian non-stationary extreme value approach to model traffic extreme conflicts across different intersections jointly for crash estimation, considering traffic volume, shock wave area, backward-moving shock wave speed, and platoon ratio. Nazir et al. (2023) adopted a Bayesian non-stationary generalized extreme value modeling approach to estimate car-following crash risk in a connected environment, considering driving behavior variables and demographic variables. Tahir and Haque (2024) proposed a non-stationary bivariate extreme value model to estimate real-time pedestrian crash risk by severity. Cavadas et al. (2020) extended bivariate non-stationary EVT models for the joint probability of head-on and rear-end crashes during passing maneuvers based on detailed trajectory data from a driving simulator. Recently, Bayesian hierarchical structure has often been combined with non-stationary model together, which aim to integrate conflicts from multiple sites for crash estimation to get a better estimation accuracy (Guo et al., 2020a; Ali et al., 2023; Zheng and Sayed, 2020; Fu et al.,2020; Arun et al., 2023). Guo et al. (2020a) proposed a Bayesian hierarchical peak over threshold approach for conflict-based before-after safety evaluation of leading pedestrian intervals, which integrates traffic conflicts from various sites and time periods. Ali et al. (2023) applied hierarchical EVT models to estimate the vehicle–pedestrian crash risk at a signal cycle level. They considered the count and speed of pedestrian and vehicles. Zheng and Sayed (2020) and Fu et al. (2020) developed bivariate and multivariate Bayesian hierarchical model for the non-stationary traffic extreme conflicts to estimate crashes, respectively.

**2.2. Threshold determination in peak over threshold method**

In the peak over threshold method, determining an appropriate threshold is crucial for accurately modeling extreme values. The threshold must be high enough to ensure the conflicts are truly extreme, yet not so high that useful conflicts are discarded. Many studies have been devoted to this topic (Scarrott and MacDonald, 2012; Coles et al., 2001; Drees et al., 2000; DuMouchel, 1983; Zheng et al., 2019b). Graphical diagnostics approach based on visual inspection of plots are widely utilized and considered the most popular method for determining thresholds. This method primarily includes the mean residual life plot (Davison and Smith, 1990), threshold stability parameters plot (Coles, 2001), and Hill plot (Drees et al., 2000). The threshold can be obtained when the mean residual life plot exhibits approximate linearity, and the threshold stability plot demonstrates approximate constancy. However, this method is subjective and may result in multiple potential thresholds, leading to significant influence on estimation outcomes (Scarrott and MacDonald, 2012). Rules of thumb is another way for threshold determination. Roth et al. (2014) chose upper 5% as threshold to estimate extreme precipitation. While Zheng et al. (2019b) used the upper 12% as the threshold to estimate crash at signalized intersection, considering the sample sizes of the collected conflict data. Obviously, the use of upper quantile as threshold varies depending on the specific problem and the sample sizes, leading to huge challenges in application of this method. Recently, researchers have developed a quantile regression approach to determine the threshold, considering it as a function of covariates (Zheng et al., 2019b; Fu and Sayed, 2023). However, selecting the appropriate quantile remains a subjective process.

To address the subjectivity and randomness in threshold determination, several research have developed numerical approaches (Behrens et al., 2004; Carreau and Bengio,2009; Zheng et al.,2014b; Northrop et al., 2017; Yue et al., 2024). Behrens et al. (2004) introduced a mixture model that combined gamma distribution and generalized Pareto distribution, utilizing all observations for inference about the threshold parameter. Carreau and Bengio (2009) proposed a hybrid Pareto distribution, which can be utilized in a mixture model, to extend the generalized Pareto distribution to the entire real axis. Zheng et al. (2014b) developed shifted Gamma-GPD distribution model to map the safety continuum and determinate the threshold. Northrop et al. (2017) utilized Bayesian model-averaging to combine inferences from multiple thresholds, thereby reducing sensitivity to the choice of a single threshold. Yue et al. (2024) proposed discontinuous, continuous, and differentiable Gamma-GPD models for the threshold determination. However, all these studies assume that the threshold in the peak over threshold method is constant, overlooking the possibility that the threshold may change with the traffic condition (such as traffic volume, road design speed), which may lead to biased estimation (Arun et al.,2023; Fu and Sayed,2023).

## 3. Model Development

In this study, a BHHM framework is proposed to fit the traffic conflicts data characterized by extremes, where the threshold for distinguishing extreme conflicts and general conflicts is taken as a model parameter which needs to be estimated. More specifically, the threshold is estimated by the proposed BHHM model where the general conflict beyond the threshold is fitted by specific distribution (i.e. normal, Cauchy, logistics, gamma, and lognormal) and the extreme conflicts below the threshold is fitted by the GPD. It should be noted that these two kinds of distributions are integrated into a unified model. Moreover, the BHHM is used to combine traffic conflict data from different sites, incorporating possible covariates that influence the non-stationary of extreme conflicts.

**3.1. Peak over threshold model**

Mathematically, let $X_1$, $X_2$, ..., $X_n$ are independently and identically distributed random observations with identical distribution function $F(x) = \Pr(X_i \leq x)$, a threshold can be set to treat variables exceeding this threshold as extreme values. This approach is commonly known as the peak over threshold method. Observations exceeding the threshold $\mu$ are considered as exceedances (i.e. extreme conflicts). As described by Pickands (1975), $F_\mu(x)$ signifies the distribution function of random variable $X_i$ exceeding the threshold $\mu$ and is denoted as

$$F_\mu(x) = \Pr\{X - \mu > x | X > \mu\} = \frac{1 - F(\mu + x)}{1 - F(\mu)}, x > 0 \qquad (1)$$

It is difficult that obtaining an exact analytical expression for the distribution in **Eq. (1)**, since $F(x)$ denotes an unknown distribution. Hence, it is necessary to construct an approximation. Assume that threshold $\mu$ is large enough, the distribution $F_\mu(x)$ can be approximated as a GPD which is represented as follows

$$G(x|\mu,\sigma,\xi) = \begin{cases} 1 - \left(1 + \xi \frac{x-\mu}{\sigma}\right)^{-1/\xi} & \xi \neq 0 \\ 1 - \exp\left(-\frac{x-\mu}{\sigma}\right) & \xi = 0 \end{cases} \qquad (2)$$

where $\mu$ refers the threshold or location parameter; $\sigma > 0$ refers the scale parameter; $-\infty < \xi < +\infty$ refers the shape parameter. $G(\cdot)$ refers the cumulative distribution function (CDF) of GPD.

In general, the above model is performed in two steps. First, the threshold $\mu$ is chosen using the mean residual life plot, the rules of thumb, or the quantile regression approach as discussed in the literature. Second, assuming that $\mu$ is known, the other parameters are estimated (Zheng et al., 2014a; Guo et al., 2020a). However, the main drawback of this idea is that there is uncertainty in the choice of the threshold $\mu$. Such subjective and arbitrary selection of the threshold can result in biased estimation outcomes.

**3.2. Hybrid structure model**

To eliminate the uncertainty and subjective in the process of threshold determination, a potential approach is to treat the threshold as a model parameter and estimate it simultaneously with other model parameters. Different with previous studies which solely utilizing observations above the threshold, all traffic conflict data is used to estimate the threshold in this study. Five types of hybrid structure models were developed: the hybrid Normal-GPD model, the hybrid Cauchy-GPD model, the hybrid Logistic-GPD model, the hybrid Gamma-GPD model, and the hybrid Lognormal-GPD model. The transition form of different distributions at the threshold is of particular importance. Our previous study investigated three scenarios: (1) the CDF is continuous while the probability density function (PDF) is unrestricted, (2) the CDF is smooth and continuous while the PDF is continuous but not smooth; and (3) both the CDF and PDF are smooth and continuous. The result found that when the CDF is continuous at the threshold while the PDF is unrestricted, the hybrid structure model achieved best performance in terms of both goodness-of-fit and crash estimation accuracy. Allowing for a discontinuous PDF at the threshold provides greater flexibility in capturing the distinct mechanisms governing general conflicts and extreme conflicts. Such flexibility ensures that the model is less constrained by assumptions of smooth transitions, thereby improving its ability to adapt to real-world traffic conflicts. As such, this study adopts this transition form to ensure the continuity of the CDF at the threshold. More details could be referred to Yue et al. (2024).

### 3.2.1. Hybrid Normal-GPD model

As shown in **Fig. 1** and **Eq. (3)**, under the assumption that observations below the threshold follow a normal distribution with parameter $\kappa$ and $\lambda$, denoted as $N(\bullet|\kappa,\lambda)$, while extreme observations exceeding the threshold conform to the generalized Pareto distribution. Therefore, the complete distribution $F$ can be expressed as

$$F(x|\kappa,\lambda,\mu,\sigma,\xi) = \begin{cases} N(x|\kappa,\lambda) & x < \mu \\ N(\mu|\kappa,\lambda) + \left[1 - N(\mu|\kappa,\lambda)\right] G(x|\mu,\sigma,\xi) & x \geq \mu \end{cases} \quad (3)$$

$$N(x|\kappa,\lambda) = \frac{1}{\lambda\sqrt{2\pi}} \int_{-\infty}^{x} \exp\left(-\frac{(t-\kappa)^2}{2\lambda^2}\right) dt \quad (4)$$

where $N(x|\kappa,\lambda)$ refers to CDF of normal distribution; $\kappa \in R$ is the location parameter of normal distribution; $\lambda > 0$ is the scale parameter of normal distribution; $G(x|\mu,\sigma,\xi)$ refers to GPD; $\mu$ is the threshold of GPD; $\sigma > 0$ is the scale parameter of GPD; $\xi \in R$ is the shape parameter of GPD.

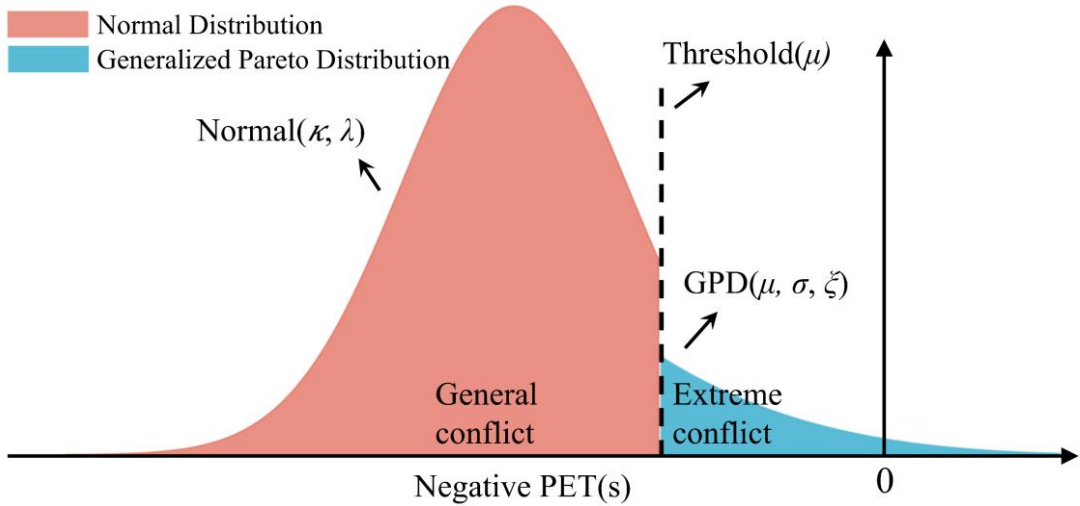

**Fig. 1.** Schematic representation of hybrid Normal-GPD model

Furthermore, the likelihood function of the hybrid Normal-GPD model can be denoted as

$$L(\boldsymbol{\theta}|x) = \begin{cases} \prod_{x<\mu} n(x|\kappa,\lambda) \prod_{x\geq\mu} \left(1-N(\mu|\kappa,\lambda)\right) \frac{1}{\sigma}\left(1+\xi\frac{x-\mu}{\sigma}\right)^{-\frac{1}{\xi}-1} & \xi \neq 0 \\ \prod_{x<\mu} n(x|\kappa,\lambda) \prod_{x\geq\mu} \left(1-N(\mu|\kappa,\lambda)\right) \frac{1}{\sigma}\exp\left(-\frac{x-\mu}{\sigma}\right) & \xi = 0 \end{cases} \quad (5)$$

$$n(x|\kappa,\lambda) = \frac{1}{\lambda\sqrt{2\pi}} \exp\left(-\frac{(x-\kappa)^2}{2\lambda^2}\right) \quad (6)$$

where $n(x|\kappa,\lambda)$ refers to PDF of normal distribution $N(x|\kappa,\lambda)$; $\boldsymbol{\theta} = (\kappa,\lambda,\mu,\xi,\sigma)$ refers to the model parameter vector.

### 3.2.2. Hybrid Cauchy-GPD model

The hybrid Cauchy-GPD model assumes that extreme observations exceeding the threshold conform to the GPD, while observations below the threshold follow the Cauchy distribution with parameters $\gamma$ and $x_0$ (see **Fig. 2**). Therefore, the complete distribution $F$ can be expressed as

$$F(x|\gamma,x_0,\mu,\sigma,\xi) = \begin{cases} C(x|\gamma,x_0) & x < \mu \\ C(\mu|\gamma,x_0) + \left[1-C(\mu|\gamma,x_0)\right]G(x|\mu,\sigma,\xi) & x \geq \mu \end{cases} \quad (7)$$

$$C(x|\gamma,x_0) = \frac{1}{2} + \frac{1}{\pi}\arctan\left(\frac{x-x_0}{\gamma}\right) \quad (8)$$

where $C(x|\gamma,x_0)$ refers to CDF of Cauchy distribution; $\gamma > 0$ is the scale parameter of Cauchy distribution; $x_0 \in \mathbb{R}$ is the location parameter of Cauchy distribution.

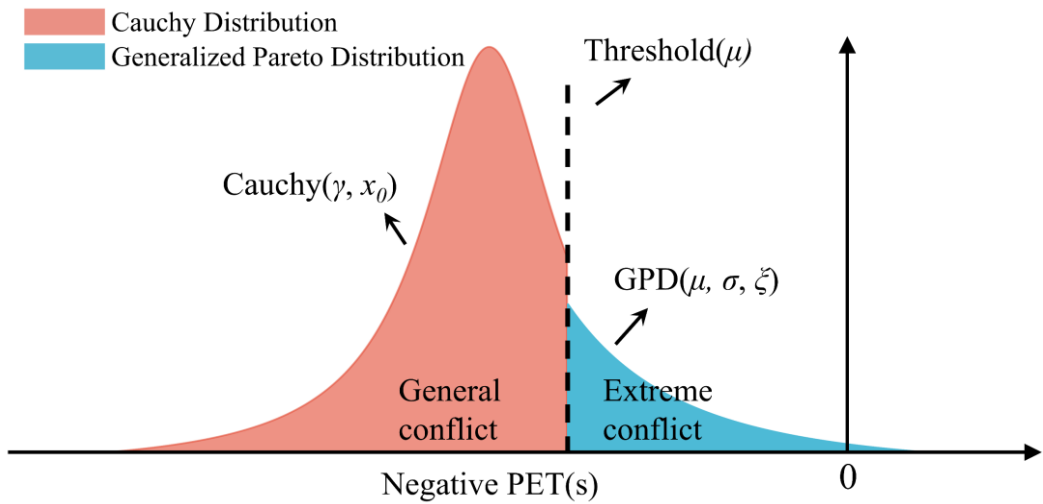

**Fig. 2.** Schematic representation of hybrid Cauchy-GPD model

Then, the likelihood function of the hybrid Cauchy-GPD model can be written as:

$$L(\boldsymbol{\theta}|x) = \begin{cases} \prod_{x<\mu} c(x|\gamma,x_0) \prod_{x\geq\mu}\left(1-C(\mu|\gamma,x_0)\right)\frac{1}{\sigma}\left(1+\xi\frac{x-\mu}{\sigma}\right)^{-\frac{1}{\xi}-1} & \xi \neq 0 \\ \prod_{x<\mu} n(x|\gamma,x_0) \prod_{x\geq\mu}\left(1-C(\mu|\gamma,x_0)\right)\frac{1}{\sigma}\exp\left(-\frac{x-\mu}{\sigma}\right) & \xi = 0 \end{cases} \quad (9)$$

$$c(x|\gamma,x_0) = \frac{1}{\pi} \cdot \frac{\gamma}{(x-x_0)^2 + \gamma^2} \quad (10)$$

where $c(x|\gamma,x_0)$ refers to PDF of Cauchy distribution $C(x|\gamma,x_0)$; $\boldsymbol{\theta}=(\gamma,x_0,\mu,\xi,\sigma)$ refers to the model parameter vector.

### 3.2.3. Hybrid Logistic-GPD model

As illustrated in **Fig. 3** and **Eq. (11)**, assuming that observations below the threshold follow a logistic distribution with parameters $\vartheta$ and $g$, denoted as $L_{ogistic}(\bullet|\vartheta,g)$, while extreme observations exceeding the threshold adhere to the GPD, the overall distribution $F$ can thus be expressed as

$$F(x|\vartheta,g,\mu,\sigma,\xi) = \begin{cases} L_{ogistic}(x|\vartheta,g) & x<\mu \\ L_{ogistic}(\mu|\vartheta,g) + \left[1-L_{ogistic}(\mu|\vartheta,g)\right]G(x|\mu,\sigma,\xi) & x\geq\mu \end{cases} \quad (11)$$

$$L_{ogistic}(x|\vartheta,g) = \frac{1}{1+\exp(-(x-\vartheta)/g)} \quad (12)$$

where $L_{ogistic}(x|\vartheta,g)$ refers to CDF of logistic distribution; $\vartheta \in R$ is the location parameter of logistic distribution; $g > 0$ is the scale parameter of logistic distribution.

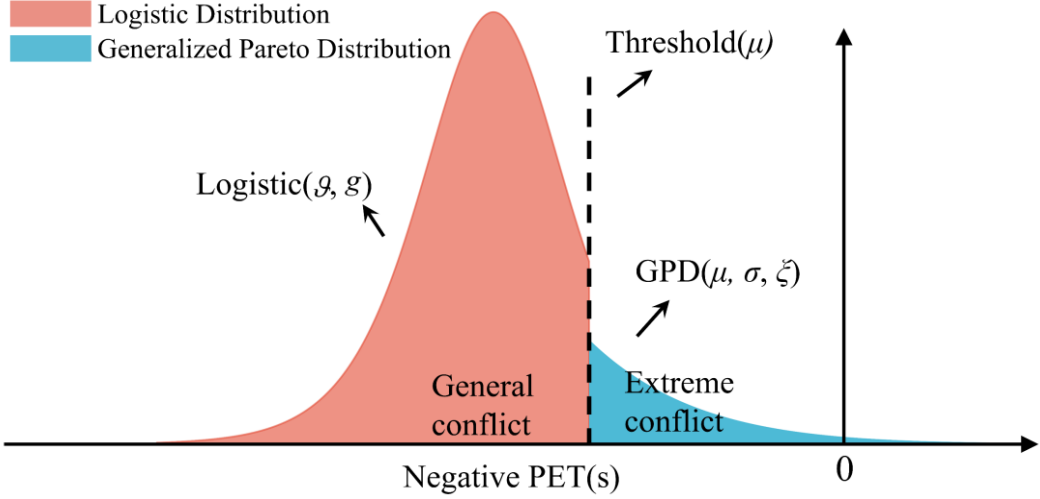

**Fig. 3.** Schematic representation of hybrid Logistic-GPD model

Additionally, the likelihood function for the hybrid Logistic-GPD model can be formulated as

$$L(\boldsymbol{\theta}|x) = \begin{cases} \prod_{x<\mu} l_{ogistic}(x|\vartheta,g) \prod_{x\geq\mu}\left(1-L_{ogistic}(\mu|\vartheta,g)\right)\frac{1}{\sigma}\left(1+\xi\frac{x-\mu}{\sigma}\right)^{-\frac{1}{\xi}-1} & \xi \neq 0 \\ \prod_{x<\mu} l_{ogistic}(x|\vartheta,g) \prod_{x\geq\mu}\left(1-L_{ogistic}(\mu|\vartheta,g)\right)\frac{1}{\sigma}\exp\left(-\frac{x-\mu}{\sigma}\right) & \xi = 0 \end{cases} \quad (13)$$

$$l_{ogistic}(x|\vartheta,g) = \frac{\exp(-(x-\vartheta)/g)}{g\left(1+\exp(-(x-\vartheta)/g)\right)^2} \quad (14)$$

where $l_{ogistic}(x|\vartheta,g)$ refers to PDF of logistic distribution $L_{ogistic}(x|\vartheta,g)$; $\boldsymbol{\theta}=(\vartheta,g,\mu,\xi,\sigma)$ refers to the model parameter vector.

### 3.2.4. Hybrid Gamma-GPD model

Since the Gamma distribution is defined over the positive domain, and this study involves modeling negative PET values, a mirrored version of the gamma distribution is employed to accommodate the data characteristics. The PDF of the standard gamma distribution is defined as:

$$z(x|p,q) = \frac{x^{p-1}q^p \exp(-qx)}{\Gamma(p)}, x > 0 \quad (15)$$

where $p > 0$ is the shape parameter of gamma distribution; $q > 0$ is the scale parameter of gamma distribution.

Correspondingly, the PDF of the mirrored gamma distribution is formulated as:

$$z(-x|p,q) = \frac{(-x)^{p-1}q^p \exp(qx)}{\Gamma(p)}, x<0 \tag{16}$$

Therefore, the hybrid Gamma-GPD model assumes that under the assumption that observations below the threshold follow a mirrored gamma distribution with parameter $p$ and $q$, while extreme observations exceeding the threshold conform to the GPD, as shown in **Fig. 4**.

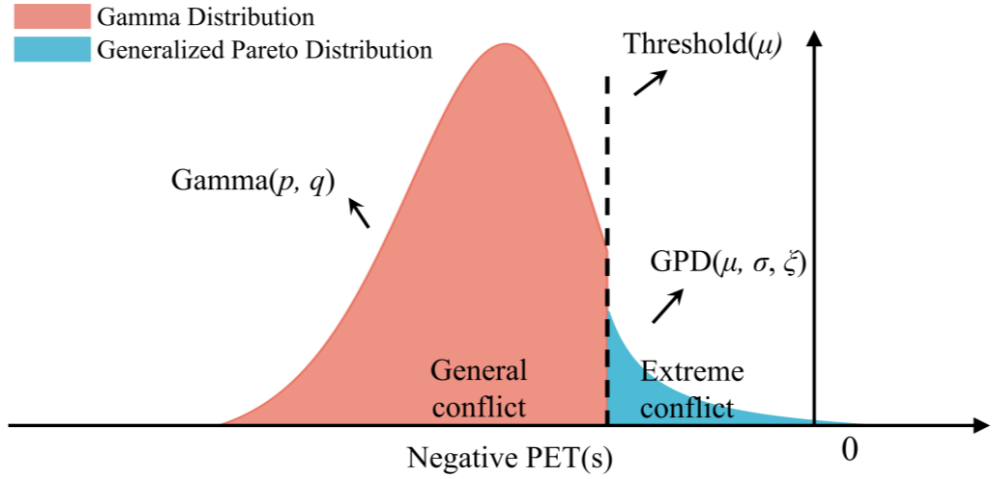

**Fig. 4.** Schematic representation of hybrid Gamma-GPD model

The complete distribution $F$ can be expressed as

$$F(x|p,q,\mu,\sigma,\xi) = \begin{cases} Z(x|p,q) & x<\mu \\ Z(\mu|p,q)+[1-Z(\mu|p,q)]\times G(x|\mu,\sigma,\xi) & x\geq \mu \end{cases} \tag{17}$$

$$Z(x|p,q) = 1 - \frac{1}{\Gamma(p)}\int_0^{-x} t^{p-1}\exp(-qt)dt \tag{18}$$

where $Z(x|p,q)$ refers to the CDF of mirrored gamma distribution. The likelihood function of the hybrid Gamma-GPD model can be written as

$$L(\boldsymbol{\theta}|x) = \begin{cases} \prod_{x<\mu} z(x|p,q) \times \prod_{x\geq\mu}(1-Z(\mu|p,q)) \times \frac{1}{\sigma} \times \left(1+\xi\times\frac{x-\mu}{\sigma}\right)^{-\frac{1}{\xi}-1} & \xi \neq 0 \\ \prod_{x<\mu} z(x|p,q) \times \prod_{x\geq\mu}(1-Z(\mu|p,q)) \times \frac{1}{\sigma} \times \exp\left(-\frac{x-\mu}{\sigma}\right) & \xi = 0 \end{cases} \tag{19}$$

where $z(x|p,q)$ refers to PDF of mirrored gamma distribution $Z(x|p,q)$; $\theta = (p,q,\mu,\xi,\sigma)$ refers to the model parameter vector.

**3.2.5. Hybrid Lognormal-GPD model**

Similar to the Gamma distribution, the lognormal distribution is also defined over the positive domain. To account for the data characteristics, a mirrored version of the lognormal distribution is used. The PDF of the standard lognormal distribution is given as

$$h(x|\omega,v) = \frac{1}{x\omega\sqrt{2\pi}} \exp\left(-\frac{(\ln(x)-v)^2}{2\omega^2}\right), x > 0 \qquad (20)$$

where $w > 0$ is the scale parameter of lognormal distribution; $v \in \mathbb{R}$ is the location parameter of lognormal distribution. Correspondingly, the PDF of the mirrored lognormal distribution is formulated as:

$$h(-x|\omega,v) = \frac{1}{-x\omega\sqrt{2\pi}} \exp\left(-\frac{(\ln(-x)-v)^2}{2\omega^2}\right), x < 0 \qquad (21)$$

Therefore, the hybrid Lognormal-GPD model assumes that under the assumption that observations below the threshold follow a lognormal distribution with parameter $w$ and $v$, denoted as $h(\cdot|w,v)$, while extreme observations exceeding the threshold follow a GPD, as shown in **Fig. 5**.

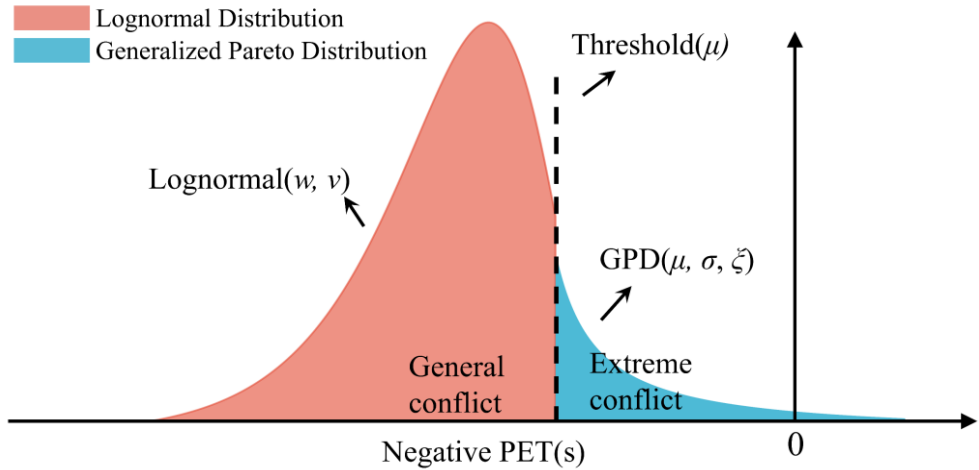

**Fig. 5.** Schematic representation of hybrid Lognormal-GPD model

Therefore, the complete distribution $F$ can be expressed as

$$F(x|w,v,\mu,\sigma,\xi) = \begin{cases} H(x|w,v) & x < \mu \\ H(\mu|w,v) + \left[1 - H(\mu|w,v)\right] \times G(x|\mu,\sigma,\xi) & x \geq \mu \end{cases} \quad (22)$$

$$H(x|w,v) = 1 - \Phi\left(\frac{\ln(-x) - v}{w}\right) \quad (23)$$

where $H(x|w,v)$ refers to the CDF of mirrored lognormal distribution; $\Phi(\bullet)$ denotes the CDF of the standard normal distribution. Then, the likelihood function of the hybrid Lognormal-GPD model can be written as

$$L(\boldsymbol{\theta}|x) = \begin{cases} \prod_{x<\mu} h(x|w,v) \times \prod_{x \geq \mu}(1 - H(\mu|w,v)) \times \frac{1}{\sigma} \times \left(1 + \xi \times \frac{x-\mu}{\sigma}\right)^{-\frac{1}{\xi}-1} & \xi \neq 0 \\ \prod_{x<\mu} h(x|w,v) \times \prod_{x \geq \mu}(1 - H(\mu|w,v)) \times \frac{1}{\sigma} \times \exp\left(-\frac{x-\mu}{\sigma}\right) & \xi = 0 \end{cases} \quad (24)$$

where $h(x|w,v)$ refers to the PDF of mirrored lognormal distribution $H(x|w,v)$; $\boldsymbol{\theta} = (w,v,\mu,\xi,\sigma)$ refers to the model parameter vector.

### 3.3. Bayesian hierarchical structure

Following the framework of our previous work (Zheng and Sayed, 2019a; Zheng et al., 2019b; Fu et al., 2020; Guo et al., 2020a; Zheng and Sayed, 2020; Fu and Sayed, 2023), this study develops a three-level Bayesian hierarchical model framework to combine traffic conflict data from different sites considering the site-specific unobserved heterogeneity, incorporating the influence of possible covariates on the non-stationary of extreme conflicts. The Bayesian hierarchical model allows for accommodating complex model structures and enables handling the small sample issues. In detail, the first layer involves modeling the observed conflict data using a hybrid parametrical model. The second layer represents a latent process underlying extreme conflicts through link functions, which establishes connections between the hybrid structure model parameters of the data layer and specific covariates. The third layer encompasses prior distributions of model parameters governing the latent process. Applying Bayes theorem within the framework of the three-layer hierarchical model, the inference for the model parameters $\Theta$ given the data $Y$ is expressed as

$$p(\Theta|Y) \propto \underbrace{p_1(Y|\theta_1)}_{\text{Data}} \underbrace{p_2(\theta_1|\theta_2)}_{\text{Process}} \underbrace{p_3(\theta_2)}_{\text{Prior}} \qquad (25)$$

where $p(\Theta|Y)$ is the posterior distribution for parameter $\Theta$; $p_j(\bullet)$ is the probability density associated with layer $j$ of the hierarchical model and depends on parameters $\theta_j$; $\theta_1$ and $\theta_2$ are the vector of parameters in data layer model and process level model, respectively. The posterior distribution of the model parameters was inferred through Markov Chain Monte Carlo simulation by utilizing Gibbs Sampling.

### 3.3.1. Data layer

At the first layer of the hierarchical model, a hybrid parametrical form is employed to model the conflict data. Specifically, specific distribution (normal, Cauchy, logistic, gamma, and lognormal) is employed to model the general conflicts, while GPD is employed to model the extreme conflicts. To adhere to the positivity constraint of the scale parameter of GPD, it is re-parameterized as $\varphi = \log \sigma$.

**(1) Hybrid Normal-GPD model**

To ensure the scale parameter $\lambda$ of the normal distribution remains positive, it is re-parameterized as $\hat{\lambda} = log(\lambda)$. Accordingly, the hybrid Normal-GPD ($\kappa, \hat{\lambda}, \mu, \xi, \sigma$) is reparametrized as Normal-GPD ($\kappa, \hat{\lambda}, \mu, \xi, \varphi$). Let $x_{ik}$ represent the $k$th observation at site $i$, where $k=1, ..., n_i$ and $i=1, ..., s$. Let $(\kappa_{ik}, \hat{\lambda}_{ik})$ and $(\mu_{ik}, \xi_{ik}, \varphi_{ik})$ be the normal distribution and GPD parameters at site $i$ for $k$th observation, respectively. The density function of Normal-GPD model is given by

$$f(x_{ik}) = \begin{cases} \dfrac{1}{\sqrt{2\pi}\exp(\hat{\lambda}_{ik})}\exp\left(-\dfrac{1}{2}\left(\dfrac{x_{ik}-\kappa_{ik}}{\exp(\hat{\lambda}_{ik})}\right)^2\right) & x_{ik} < \mu_{ik} \\ \left(1-\Phi\left(\dfrac{\mu_{ik}-\kappa_{ik}}{\exp(\hat{\lambda}_{ik})}\right)\right)\times\dfrac{1}{\exp(\varphi_{ik})}\left(1+\xi_{ik}\dfrac{x_{ik}-\mu_{ik}}{\exp(\varphi_{ik})}\right)^{-\dfrac{1}{\xi_{ik}}-1} & x_{ik} \geq \mu_{ik} \end{cases} \quad (26)$$

The first term in the right-hand side in **Eq. (25)** is given by the likelihood function

$$p_1(Y|\theta_1) = \prod_{i=1}^{s}\prod_{x_{ik}<\mu_{ik}} \dfrac{1}{\sqrt{2\pi}\times\exp(\hat{\lambda}_{ik})}\times\exp\left(-\dfrac{1}{2}\left(\dfrac{x_{ik}-\kappa_{ik}}{\exp(\hat{\lambda}_{ik})}\right)^2\right)$$
$$\times\prod_{x_{ik}\geq\mu_{ik}}\left(\left(1-\Phi\left(\dfrac{\mu_{ik}-\kappa_{ik}}{\exp(\hat{\lambda}_{ik})}\right)\right)\times\dfrac{1}{\exp(\varphi_{ik})}\times\left(1+\xi_{ik}\times\dfrac{x_{ik}-\mu_{ik}}{\exp(\varphi_{ik})}\right)^{-\dfrac{1}{\xi_{ik}}-1}\right) \quad (27)$$

where $\theta_1 = \left[\kappa,\hat{\lambda},\mu,\xi,\varphi\right]^T$.

**(2) Hybrid Cauchy-GPD model**

The scale parameter $\gamma$ of the Cauchy distribution is re-parameterized as $\hat{\gamma} = log(\gamma)$ to maintain its positivity. As a result, the hybrid Cauchy-GPD ($\gamma$, $x_0$, $\mu$, $\xi$, $\sigma$) is reparametrized as Cauchy-GPD ($\hat{\gamma}$, $x_0$, $\mu$, $\xi$, $\varphi$). Let $(\hat{\gamma}_{ik}, x_{0ik})$ and $(\mu_{ik}, \xi_{ik}, \varphi_{ik})$ be the Cauchy distribution and GPD parameters at site $i$ for $k$th observation, respectively. The density function of Cauchy-GPD model is expressed as

$$f(x_{ik}) =$$
$$\begin{cases} \dfrac{1}{\pi}\times\dfrac{\exp(\hat{\gamma}_{ik})}{(x_{ik}-x_{0ik})^2+(\exp(\hat{\gamma}_{ik}))^2} & x_{ik} < \mu_{ik} \\ \left(\dfrac{1}{2}-\dfrac{1}{\pi}\times\arctan\left(\dfrac{\mu_{ik}-x_{0ik}}{\exp(\hat{\gamma}_{ik})}\right)\right)\times\dfrac{1}{\exp(\varphi_{ik})}\left(1+\xi_{ik}\times\dfrac{x_{ik}-\mu_{ik}}{\exp(\varphi_{ik})}\right)^{-\dfrac{1}{\xi_{ik}}-1} & x_{ik} \geq \mu_{ik} \end{cases} \quad (28)$$

The first term in the right-hand side in **Eq. (25)** is given by the likelihood function

$$p_1(Y|\theta_1) =$$

$$\prod_{i=1}^{s} \prod_{x_{ik} < \mu_{ik}} \frac{1}{\pi} \times \frac{\exp(\hat{\gamma}_{ik})}{(x_{ik} - x_{0ik})^2 + (\exp(\hat{\gamma}_{ik}))^2} \qquad (29)$$

$$\times \prod_{x_{ik} \geq \mu_{ik}} \left( \left( \frac{1}{2} - \frac{1}{\pi} \arctan\left( \frac{\mu_{ik} - x_{0ik}}{\exp(\hat{\gamma}_{ik})} \right) \right) \times \frac{1}{\exp(\varphi_{ik})} \times \left( 1 + \xi_{ik} \times \frac{x_{ik} - \mu_{ik}}{\exp(\varphi_{ik})} \right)^{-\frac{1}{\xi_{ik}} - 1} \right)$$

where $\theta_1 = [\kappa, \hat{\gamma}, \mu, \xi, \varphi]^T$.

**(3) Hybrid Logistic-GPD model**

To guarantee the positivity of the scale parameter $g$ in the logistic distribution, it is reparameterized as $\hat{g} = \log(g)$. Consequently, the hybrid Logistic-GPD ($\vartheta$, $g$, $\mu$, $\xi$, $\sigma$) is reparametrized as Logistic-GPD ($\vartheta$, $\hat{g}$, $\mu$, $\xi$, $\varphi$). Let ($\vartheta_{ik}$, $\hat{g}_{ik}$) and ($\mu_{ik}$, $\xi_{ik}$, $\varphi_{ik}$) be the logistic distribution and GPD parameters at site $i$ for $k$th observation, respectively. The density function of Logistic-GPD model is given by

$$f(x_{ik}) = \begin{cases} \dfrac{\exp(-(x_{ik} - \vartheta_{ik})/\exp(\hat{g}_{ik}))}{\exp(\hat{g}_{ik}) \times (1 + \exp(-(x_{ik} - \vartheta_{ik})/\exp(\hat{g}_{ik})))^2} & x_{ik} < \mu_{ik} \\ \dfrac{\exp(-(\mu_{ik} - \vartheta_{ik})/\exp(\hat{g}_{ik}))}{1 + \exp(-(\mu_{ik} - \vartheta_{ik})/\exp(\hat{g}_{ik}))} \times \dfrac{1}{\exp(\varphi_{ik})} \times \left(1 + \xi_{ik} \dfrac{x_{ik} - \mu_{ik}}{\exp(\varphi_{ik})}\right)^{-\frac{1}{\xi_{ik}} - 1} & x_{ik} \geq \mu_{ik} \end{cases} \qquad (30)$$

The first term in the right-hand side in **Eq. (25)** is given by the likelihood function

$$p_1(Y|\theta_1) = \prod_{i=1}^{s} \prod_{x_{ik} < \mu_{ik}} \frac{\exp(-(x_{ik} - \vartheta_{ik})/\exp(\hat{g}_{ik}))}{\exp(\hat{g}_{ik}) \times (1 + \exp(-(x_{ik} - \vartheta_{ik})/\exp(\hat{g}_{ik})))^2} \qquad (31)$$

$$\times \prod_{x_{ik} \geq \mu_{ik}} \left( \frac{\exp(-(\mu_{ik} - \vartheta_{ik})/\exp(\hat{g}_{ik}))}{1 + \exp(-(\mu_{ik} - \vartheta_{ik})/\exp(\hat{g}_{ik}))} \times \frac{1}{\exp(\varphi_{ik})} \times \left(1 + \xi_{ik} \frac{x_{ik} - \mu_{ik}}{\exp(\varphi_{ik})}\right)^{-\frac{1}{\xi_{ik}} - 1} \right)$$

where $\theta_1 = [\vartheta, \hat{g}, \mu, \xi, \varphi]^T$.

**(4) Hybrid Gamma-GPD model**

To ensure the positivity of the shape parameter $p$ and the scale parameter $q$ in the gamma distribution, they are re-parameterized as $\hat{p} = log(p)$ and $\hat{q} = log(q)$. Consequently, the hybrid Gamma-GPD ($p, q, \mu, \xi, \sigma$) is reparametrized as Gamma-GPD ($\hat{p}, \hat{q}, \mu, \xi, \varphi$). Let ($\hat{p}_{ik}, \hat{q}_{ik}$) and ($\mu_{ik}, \xi_{ik}, \varphi_{ik}$) be the gamma distribution and GPD parameters at site $i$ for $k$th observation, respectively. The density function of Gamma-GPD model is expressed as

$$f(x_{ik}) = \begin{cases} \dfrac{(-x_{ik})^{\hat{p}_{ik}-1} \times \hat{q}_{ik}^{\hat{p}_{ik}} \times \exp(\hat{q}_{ik} x_{ik})}{\Gamma(\hat{p}_{ik})} & x_{ik} < \mu_{ik} \\ Z(\mu_{ik} | \hat{p}_{ik}, \hat{q}_{ik}) \times \dfrac{1}{\exp(\varphi_{ik})} \times \left(1 + \xi_{ik} \dfrac{x_{ik} - \mu_{ik}}{\exp(\varphi_{ik})}\right)^{-\frac{1}{\xi_{ik}}-1} & x_{ik} \geq \mu_{ik} \end{cases} \quad (32)$$

The first term in the right-hand side in **Eq. (25)** is given by the likelihood function

$$p_1(Y|\theta_1) = \prod_{i=1}^{s} \prod_{x_{ik} < \mu_{ik}} \dfrac{(-x_{ik})^{\hat{p}_{ik}-1} \times \hat{q}_{ik}^{\hat{p}_{ik}} \times \exp(\hat{q}_{ik} x_{ik})}{\Gamma(\hat{p}_{ik})}$$
$$\times \prod_{x_{ik} \geq \mu_{ik}} \left( Z(\mu_{ik} | \hat{p}_{ik}, \hat{q}_{ik}) \times \dfrac{1}{\exp(\varphi_{ik})} \times \left(1 + \xi_{ik} \dfrac{x_{ik} - \mu_{ik}}{\exp(\varphi_{ik})}\right)^{-\frac{1}{\xi_{ik}}-1} \right) \quad (33)$$

where $\theta_1 = \left[\hat{p}, \hat{q}, \mu, \xi, \varphi\right]^T$.

**(5) Hybrid Lognormal-GPD model**

To ensure the scale parameter $w$ of the lognormal distribution remains positive, it is re-parameterized as $\hat{w} = log(w)$. Therefore, the hybrid Lognormal-GPD model, initially formulated as Lognormal-GPD ($w, v, \mu, \xi, \sigma$), is reparametrized as Lognormal-GPD ($\hat{w}, v, \mu, \xi, \varphi$). Let $x_{ik}$ represent the $k$th observation at site $i$, where $k=1, ..., n_i$ and $i=1, ..., s$. The parameters corresponding to the lognormal distribution and GPD for the $k$th observation at site $i$ are represented by ($\hat{w}_{ik}, v_{ik}$) and ($\mu_{ik}, \xi_{ik}, \varphi_{ik}$), respectively. The density function of Lognormal-GPD model is expressed as

$$f(x_{ik}) = \begin{cases} \dfrac{1}{-x_{ik} \times \exp(\hat{\omega}_{ik}) \times \sqrt{2\pi}} \times \exp\left(-\dfrac{\left(\ln(-x_{ik})-v_{ik}\right)^2}{2\times\left(\exp(\hat{\omega}_{ik})\right)^2}\right) & x_{ik} < \mu_{ik} \\[2ex] \Phi\left(\dfrac{\ln(-\mu_{ik})-v_{ik}}{\exp(\hat{\omega}_{ik})}\right) \times \dfrac{1}{\exp(\varphi_{ik})} \times \left(1+\xi_{ik}\dfrac{x_{ik}-\mu_{ik}}{\exp(\varphi_{ik})}\right)^{-\frac{1}{\xi_{ik}}-1} & x_{ik} \geq \mu_{ik} \end{cases} \quad (34)$$

The first term in the right-hand side in **Eq. (25)** is given by the likelihood function

$$p_1(Y|\theta_1) = \prod_{i=1}^{s} \prod_{x_{ik}<\mu_{ik}} \dfrac{1}{-x_{ik} \times \exp(\hat{\omega}_{ik}) \times \sqrt{2\pi}} \times \exp\left(-\dfrac{\left(\ln(-x_{ik})-v_{ik}\right)^2}{2\times\left(\exp(\hat{\omega}_{ik})\right)^2}\right)$$
$$\times \prod_{x_{ik}\geq\mu_{ik}} \left(\Phi\left(\dfrac{\ln(-\mu_{ik})-v_{ik}}{\exp(\hat{\omega}_{ik})}\right) \times \dfrac{1}{\exp(\varphi_{ik})} \times \left(1+\xi_{ik}\dfrac{x_{ik}-\mu_{ik}}{\exp(\varphi_{ik})}\right)^{-\frac{1}{\xi_{ik}}-1}\right) \quad (35)$$

where $\theta_1 = [\hat{w}, v, \mu, \xi, \varphi]^T$.

### 3.3.2. Process layer

At the second layer, a latent Gaussian process is used to model the traffic conflicts by constructing a structure that related the parameters of the data layer to specific covariates with identity link functions. The link functions are used to relate both $\mu_i$ and $\varphi_i$ to specific covariates, addressing the non-stationary nature of extreme conflicts. Additionally, this study connects the parameters of the specific distributions to covariates to account for the non-stationary of general traffic conflicts. Given that it is challenging to precisely estimate the shape parameter $\xi$ of GPD (Coles et al., 2001), it is not feasible to link $\xi$ as a smooth function of certain covariates (Cooley et al., 2006). Following the approach of Zheng and Sayed (2019a), the shape parameter $\xi$ is assumed to be stable and unaffected by covariates in this study.

**(1) Hybrid Normal-GPD model**

In this study, it is assumed that the parameters ($\kappa_i$ and $\hat{\lambda}_i$) of the normal distribution are linked to specific covariates. The parameters of the hybrid Normal-GPD model in the process layer can be expressed as

$$\begin{cases} \mu_i = \beta_{\mu_0} + \beta_\mu X \\ \varphi_i = \beta_{\varphi_0} + \beta_\varphi X \\ \xi_i = \beta_{\xi_0} \\ \kappa_i = \beta_{\kappa_0} + \beta_\kappa X \\ \hat{\lambda}_i = \beta_{\lambda_0} + \beta_\lambda X \end{cases} \qquad (36)$$

where $X$ is the vector of covariates; $\beta_{\mu_0}, \beta_{\varphi_0}, \beta_{\xi_0}, \beta_{\kappa_0}, \beta_{\lambda_0}$ are intercepts; $\beta_\mu, \beta_\varphi, \beta_\kappa, \beta_\lambda$ are the vectors of estimated parameters of covariates. Besides, to capture the additional unobserved heterogeneity, random error terms are introduced as follows

$$\begin{cases} \mu_i = \beta_{\mu_0} + \beta_\mu X + \varepsilon_{\mu i} \\ \varphi_i = \beta_{\varphi_0} + \beta_\varphi X + \varepsilon_{\varphi i} \\ \xi_i = \beta_{\xi_0} + \varepsilon_{\xi i} \\ \kappa_i = \beta_{\kappa_0} + \beta_\kappa X + \varepsilon_{\kappa i} \\ \hat{\lambda}_i = \beta_{\lambda_0} + \beta_\lambda X + \varepsilon_{\lambda i} \end{cases} \qquad (37)$$

where $\varepsilon_{\mu i}, \varepsilon_{\varphi i}, \varepsilon_{\xi i}, \varepsilon_{\kappa i}, \varepsilon_{\lambda i}$ are random effects representing between-site variation. It should be noted that the random effects are consistent in the same site but vary in different sites. Therefore, the model for the process layer can be expressed in the following form of a random intercept.

$$\begin{cases} \mu_i = \beta_{\mu_0 i} + \beta_\mu X \\ \varphi_i = \beta_{\varphi_0 i} + \beta_\varphi X \\ \xi_i = \beta_{\xi_0 i} \\ \kappa_i = \beta_{\kappa_0 i} + \beta_\kappa X \\ \hat{\lambda}_i = \beta_{\lambda_0 i} + \beta_\lambda X \end{cases} \qquad (38)$$

where $\beta_{\mu_0 i}, \beta_{\varphi_0 i}, \beta_{\xi_0 i}, \beta_{\kappa_0 i}, \beta_{\lambda_0 i}$ are site-specific random intercepts. The random intercepts are modeled as a zero mean Gaussian process, which are $\beta_{\mu_0 i} \sim N(0, \delta_\mu^2)$, $\beta_{\varphi_0 i} \sim N(0, \delta_\varphi^2)$, $\beta_{\xi_0 i} \sim N(0, \delta_\xi^2)$, $\beta_{\kappa_0 i} \sim N(0, \delta_\kappa^2)$, $\beta_{\lambda_0 i} \sim N(0, \delta_\lambda^2)$.

With the model parameters specified above, the "process" piece of **Eq. (25)** is given by

$$p_2(\theta_1|\theta_2) = \left(\frac{1}{\sqrt{2\pi\delta_\kappa^2}}\exp\left[-\frac{(\kappa-\kappa_i)^2}{2\delta_\kappa^2}\right] \times \frac{1}{\sqrt{2\pi\delta_\lambda^2}}\exp\left[-\frac{(\hat{\lambda}-\hat{\lambda}_i)^2}{2\delta_\lambda^2}\right]\right)$$

$$\times \frac{1}{\sqrt{2\pi\delta_\mu^2}}\exp\left[-\frac{(\mu-\mu_i)^2}{2\delta_\mu^2}\right] \times \frac{1}{\sqrt{2\pi\delta_\varphi^2}}\exp\left[-\frac{(\varphi-\varphi_i)^2}{2\delta_\varphi^2}\right] \quad (39)$$

$$\times \frac{1}{\sqrt{2\pi\delta_\xi^2}}\exp\left[-\frac{(\xi-\xi_i)^2}{2\delta_\xi^2}\right] \times p_\xi(\xi|\theta_\xi)$$

where the density function $p_\xi$ comes from the prior distribution chose for the shape parameter $\xi$ with parameters $\theta_\xi$, and $\theta_2 = [\boldsymbol{\beta}_\mu, \boldsymbol{\beta}_\varphi, \boldsymbol{\beta}_\kappa, \boldsymbol{\beta}_\lambda, \beta_{\mu_0 i}, \beta_{\varphi_0 i}, \beta_{\xi_0 i}, \beta_{\kappa_0 i}, \beta_{\lambda_0 i}]^T$.

**(2) Hybrid Cauchy-GPD model**

Similar with the hybrid Normal-GPD model, the hybrid Cauchy-GPD model for the process layer can be written as

$$\begin{cases} \mu_i = \beta_{\mu_0 i} + \boldsymbol{\beta}_\mu X \\ \varphi_i = \beta_{\varphi_0 i} + \boldsymbol{\beta}_\varphi X \\ \xi_i = \beta_{\xi_0 i} \\ \hat{\gamma}_i = \beta_{\gamma_0 i} + \boldsymbol{\beta}_\gamma X \\ x_{0i} = \beta_{x_{00} i} + \boldsymbol{\beta}_{x_0} X \end{cases} \quad (40)$$

where $\beta_{\mu_0 i}, \beta_{\varphi_0 i}, \beta_{\xi_0 i}, \beta_{\gamma_0 i}, \beta_{x_{00} i}$ are site-specific random intercepts. The random intercepts are modeled as a zero mean Gaussian process, which are $\beta_{\mu_0 i} \sim N(0, \delta_\mu^2)$, $\beta_{\varphi_0 i} \sim N(0, \delta_\varphi^2)$, $\beta_{\xi_0 i} \sim N(0, \delta_\xi^2)$, $\beta_{\gamma_0 i} \sim N(0, \delta_\gamma^2)$, $\beta_{x_{00} i} \sim N(0, \delta_{x_0}^2)$.

The "process" piece of **Eq. (25)** is given by

$$p_2(\theta_1|\theta_2) = \frac{1}{\sqrt{2\pi\delta_{x_0}^2}}\exp\left[-\frac{(x_0-x_{0i})^2}{2\delta_{x_0}^2}\right] \times \frac{1}{\sqrt{2\pi\delta_\gamma^2}}\exp\left[-\frac{(\hat{\gamma}-\hat{\gamma}_i)^2}{2\delta_\gamma^2}\right]$$

$$\times \frac{1}{\sqrt{2\pi\delta_\mu^2}}\exp\left[-\frac{(\mu-\mu_i)^2}{2\delta_\mu^2}\right] \times \frac{1}{\sqrt{2\pi\delta_\varphi^2}}\exp\left[-\frac{(\varphi-\varphi_i)^2}{2\delta_\varphi^2}\right] \quad (41)$$

$$\times \frac{1}{\sqrt{2\pi\delta_\xi^2}}\exp\left[-\frac{(\xi-\xi_i)^2}{2\delta_\xi^2}\right] \times p_\xi(\xi|\theta_\xi)$$

where $\boldsymbol{\theta_2} = \left[ \boldsymbol{\beta_\mu}, \boldsymbol{\beta_\varphi}, \boldsymbol{\beta_\gamma}, \boldsymbol{\beta_{x_0}}, \beta_{\mu_0 i}, \beta_{\varphi_0 i}, \beta_{\xi_0 i}, \beta_{\gamma_0 i}, \beta_{x_{00} i} \right]^T$.

**(3) Hybrid Logistic-GPD model**

As for the hybrid Logistic-GPD model, the process layer can be expressed as

$$\begin{cases} \mu_i = \beta_{\mu_0 i} + \boldsymbol{\beta_\mu} X \\ \varphi_i = \beta_{\varphi_0 i} + \boldsymbol{\beta_\varphi} X \\ \xi_i = \beta_{\xi_0 i} \\ \kappa_i = \beta_{\kappa_0 i} + \boldsymbol{\beta_\kappa} X \\ \hat{g}_i = \beta_{g_0 i} + \boldsymbol{\beta_g} X \end{cases} \quad (42)$$

where $\beta_{\mu_0 i}, \beta_{\varphi_0 i}, \beta_{\xi_0 i}, \beta_{\vartheta_0 i}, \beta_{g_0 i}$ are site-specific random intercepts. The random intercepts are modeled as a zero mean Gaussian process, which are $\beta_{\mu_0 i} \sim N(0, \delta_\mu^2)$, $\beta_{\varphi_0 i} \sim N(0, \delta_\varphi^2)$, $\beta_{\xi_0 i} \sim N(0, \delta_\xi^2)$, $\beta_{\vartheta_0 i} \sim N(0, \delta_\vartheta^2)$, $\beta_{g_0 i} \sim N(0, \delta_g^2)$.

The "process" piece of **Eq. (25)** is given by

$$\begin{aligned} p_2(\boldsymbol{\theta_1} | \boldsymbol{\theta_2}) &= \frac{1}{\sqrt{2\pi\delta_\vartheta^2}} \exp\left[-\frac{(\vartheta - \vartheta_i)^2}{2\delta_\vartheta^2}\right] \times \frac{1}{\sqrt{2\pi\delta_g^2}} \exp\left[-\frac{(\hat{g} - \hat{g}_i)^2}{2\delta_g^2}\right] \\ &\times \frac{1}{\sqrt{2\pi\delta_\mu^2}} \exp\left[-\frac{(\mu - \mu_i)^2}{2\delta_\mu^2}\right] \times \frac{1}{\sqrt{2\pi\delta_\varphi^2}} \exp\left[-\frac{(\varphi - \varphi_i)^2}{2\delta_\varphi^2}\right] \\ &\times \frac{1}{\sqrt{2\pi\delta_\xi^2}} \exp\left[-\frac{(\xi - \xi_i)^2}{2\delta_\xi^2}\right] \times p_\xi(\xi | \theta_\xi) \end{aligned} \quad (43)$$

where $\boldsymbol{\theta_2} = \left[ \boldsymbol{\beta_\mu}, \boldsymbol{\beta_\varphi}, \boldsymbol{\beta_\vartheta}, \boldsymbol{\beta_g}, \beta_{\mu_0 i}, \beta_{\varphi_0 i}, \beta_{\xi_0 i}, \beta_{\vartheta_0}, \beta_{g_0} \right]^T$.

**(4) Hybrid Gamma-GPD model**

As for the hybrid Gamma-GPD model, the process layer is given by

$$\begin{cases} \mu_i = \beta_{\mu_0 i} + \beta_\mu X \\ \varphi_i = \beta_{\varphi_0 i} + \beta_\varphi X \\ \xi_i = \beta_{\xi_0 i} \\ \hat{p}_i = \beta_{p_0 i} + \beta_p X \\ \hat{q}_i = \beta_{q_0 i} + \beta_q X \end{cases} \tag{44}$$

where $\beta_{\mu_0 i}, \beta_{\varphi_0 i}, \beta_{\xi_0 i}, \beta_{p_0 i}, \beta_{q_0 i}$ are site-specific random intercepts. The random intercepts are modeled as a zero mean Gaussian process, which are $\beta_{\mu_0 i} \sim N(0, \delta_\mu^2)$, $\beta_{\varphi_0 i} \sim N(0, \delta_\varphi^2)$, $\beta_{\xi_0 i} \sim N(0, \delta_\xi^2)$, $\beta_{p_0 i} \sim N(0, \delta_p^2)$, $\beta_{q_0 i} \sim N(0, \delta_q^2)$.

The "process" piece of **Eq. (25)** is given by

$$p_2(\theta_1|\theta_2) = \frac{1}{\sqrt{2\pi\delta_p^2}} \exp\left[-\frac{(\hat{p}-\hat{p}_i)^2}{2\delta_p^2}\right] \times \frac{1}{\sqrt{2\pi\delta_q^2}} \exp\left[-\frac{(\hat{q}-\hat{q}_i)^2}{2\delta_q^2}\right]$$
$$\times \frac{1}{\sqrt{2\pi\delta_\mu^2}} \exp\left[-\frac{(\mu-\mu_i)^2}{2\delta_\mu^2}\right] \times \frac{1}{\sqrt{2\pi\delta_\varphi^2}} \exp\left[-\frac{(\varphi-\varphi_i)^2}{2\delta_\varphi^2}\right] \tag{45}$$
$$\times \frac{1}{\sqrt{2\pi\delta_\xi^2}} \exp\left[-\frac{(\xi-\xi_i)^2}{2\delta_\xi^2}\right] \times p_\xi(\xi|\theta_\xi)$$

where $\theta_2 = \left[\beta_\mu, \beta_\varphi, \beta_p, \beta_q, \beta_{\mu_0 i}, \beta_{\varphi_0 i}, \beta_{\xi_0 i}, \beta_{p_0}, \beta_{q_0}\right]^T$.

**(5) Hybrid Lognormal-GPD model**

The hybrid Lognormal-GPD model is similar with above four types of hybrid GPD models, the process layer is can be written as

$$\begin{cases} \mu_i = \beta_{\mu_0 i} + \beta_\mu X \\ \varphi_i = \beta_{\varphi_0 i} + \beta_\varphi X \\ \xi_i = \beta_{\xi_0 i} \\ \hat{w}_i = \beta_{w_0 i} + \beta_w X \\ v_i = \beta_{v_0 i} + \beta_v X \end{cases} \tag{46}$$

where $\beta_{\mu_0 i}, \beta_{\varphi_0 i}, \beta_{\xi_0 i}, \beta_{w_0 i}, \beta_{v_0 i}$ are site-specific random intercepts. The random intercepts are modeled as a zero mean Gaussian process, which are $\beta_{\mu_0 i} \sim N(0, \delta_\mu^2)$,

$$\beta_{\varphi_0 i} \sim N(0,\delta_\varphi^2), \beta_{\xi_0 i} \sim N(0,\delta_\xi^2), \beta_{w_0 i} \sim N(0,\delta_w^2), \beta_{v_0 i} \sim N(0,\delta_v^2).$$

The "process" piece of **Eq. (25)** is given by

$$
\begin{aligned}
p_2(\boldsymbol{\theta_1}|\boldsymbol{\theta_2}) = & \frac{1}{\sqrt{2\pi\delta_w^2}}\exp\left[-\frac{(\hat{w}-\hat{w}_i)^2}{2\delta_w^2}\right] \times \frac{1}{\sqrt{2\pi\delta_v^2}}\exp\left[-\frac{(v-v_i)^2}{2\delta_v^2}\right] \\
& \times \frac{1}{\sqrt{2\pi\delta_\mu^2}}\exp\left[-\frac{(\mu-\mu_i)^2}{2\delta_\mu^2}\right] \times \frac{1}{\sqrt{2\pi\delta_\varphi^2}}\exp\left[-\frac{(\varphi-\varphi_i)^2}{2\delta_\varphi^2}\right] \\
& \times \frac{1}{\sqrt{2\pi\delta_\xi^2}}\exp\left[-\frac{(\xi-\xi_i)^2}{2\delta_\xi^2}\right] \times p_\xi(\xi|\boldsymbol{\theta_\xi})
\end{aligned}
\qquad (47)
$$

where $\boldsymbol{\theta_2} = \left[\boldsymbol{\beta_\mu}, \boldsymbol{\beta_\varphi}, \boldsymbol{\beta_w}, \boldsymbol{\beta_v}, \beta_{\mu_0 i}, \beta_{\varphi_0 i}, \beta_{\xi_0 i}, \beta_{w_0 i}, \beta_{v_0 i}\right]^T$.

### 3.3.3. Prior layer

At the prior level, priors are assigned to the hyper-parameters to characterize the latent process in the process layer. It is assumed that each parameter is independent of the others. Given the absence of available prior information on how the model parameters are related with covariates, uninformative priors for parameters (e.g. $\left(\boldsymbol{\beta_\mu}, \boldsymbol{\beta_\varphi}, \boldsymbol{\beta_\kappa}, \boldsymbol{\beta_\lambda}, \beta_{\kappa 0}, \beta_{\lambda 0}, \beta_{\mu 0}, \beta_{\varphi 0}\right)$ ). They are hypothesized to follow a normal distribution with a mean of zero and a large variance, i.e., $N(0,10^6)$. It is essential to recognize that the GPD is highly sensitive to its shape parameter. Therefore, an improperly defined prior could result in convergence issues during model estimation. Based on previous road safety studies (Guo et al. 2020b), the estimated shape parameters typically fall within the range of (-1.0, 1.0). Hence, priors for the shape parameters are assumed to follow a uniform distribution, i.e., Unif (-1.0,1.0).

With the priors specified above, the "prior" part of **Eq. (25)** is given by

**(1) Hybrid Normal-GPD model**

$$
\begin{aligned}
p_3(\boldsymbol{\theta_2}) = & p_{\beta_\mu}(\boldsymbol{\beta_\mu}) \times p_{\beta_\varphi}(\boldsymbol{\beta_\varphi}) \times p_{\beta_\kappa}(\boldsymbol{\beta_\kappa}) \times p_{\beta_\lambda}(\boldsymbol{\beta_\lambda}) \\
& \times p_{\beta_{\kappa 0}}(\beta_{\kappa 0}) \times p_{\beta_{\lambda 0}}(\beta_{\lambda 0}) \times p_{\beta_{\mu 0}}(\beta_{\mu 0}) \times p_{\beta_{\varphi 0}}(\beta_{\varphi 0}) \times p_{\beta_{\xi 0}}(\beta_{\xi 0})
\end{aligned}
\qquad (48)
$$

**(2) Hybrid Cauchy-GPD model**

$$p_3(\boldsymbol{\theta}_2) = p_{\beta_\mu}(\boldsymbol{\beta_\mu}) \times p_{\beta_\varphi}(\boldsymbol{\beta_\varphi}) \times p_{\beta_{x_o}}(\boldsymbol{\beta_{x_o}}) \times p_{\beta_\gamma}(\boldsymbol{\beta_\gamma})$$
$$\times p_{\beta_{x_00}}(\beta_{x_00}) \times p_{\beta_{\gamma 0}}(\beta_{\gamma 0}) \times p_{\beta_{\mu 0}}(\beta_{\mu 0}) \times p_{\beta_{\varphi 0}}(\beta_{\varphi 0}) \times p_{\beta_{\xi 0}}(\beta_{\xi 0})$$
(49)

**(3) Hybrid Logistic-GPD model**

$$p_3(\boldsymbol{\theta}_2) = p_{\beta_\mu}(\boldsymbol{\beta_\mu}) \times p_{\beta_\varphi}(\boldsymbol{\beta_\varphi}) \times p_{\beta_\vartheta}(\boldsymbol{\beta_\vartheta}) \times p_{\beta_g}(\boldsymbol{\beta_g})$$
$$\times p_{\beta_{\vartheta 0}}(\beta_{\vartheta 0}) \times p_{\beta_{g 0}}(\beta_{g 0}) \times p_{\beta_{\mu 0}}(\beta_{\mu 0}) \times p_{\beta_{\varphi 0}}(\beta_{\varphi 0}) \times p_{\beta_{\xi 0}}(\beta_{\xi 0})$$
(50)

**(4) Hybrid Gamma-GPD model**

$$p_3(\boldsymbol{\theta}_2) = p_{\beta_\mu}(\boldsymbol{\beta_\mu}) \times p_{\beta_\varphi}(\boldsymbol{\beta_\varphi}) \times p_{\beta_p}(\boldsymbol{\beta_p}) \times p_{\beta_q}(\boldsymbol{\beta_q})$$
$$\times p_{\beta_{p0}}(\beta_{p0}) \times p_{\beta_{q0}}(\beta_{q0}) \times p_{\beta_{\mu 0}}(\beta_{\mu 0}) \times p_{\beta_{\varphi 0}}(\beta_{\varphi 0}) \times p_{\beta_{\xi 0}}(\beta_{\xi 0})$$
(51)

**(5) Hybrid Lognormal-GPD model**

$$p_3(\boldsymbol{\theta}_2) = p_{\beta_\mu}(\boldsymbol{\beta_\mu}) \times p_{\beta_\varphi}(\boldsymbol{\beta_\varphi}) \times p_{\beta_w}(\boldsymbol{\beta_w}) \times p_{\beta_v}(\boldsymbol{\beta_v})$$
$$\times p_{\beta_{w0}}(\beta_{w0}) \times p_{\beta_{v0}}(\beta_{v0}) \times p_{\beta_{\mu 0}}(\beta_{\mu 0}) \times p_{\beta_{\varphi 0}}(\beta_{\varphi 0}) \times p_{\beta_{\xi 0}}(\beta_{\xi 0})$$
(52)

### 3.4. Model comparison

Several non-stationary BHHM models were developed. These models are compared based on their statistical fit. In the Bayesian framework, the Deviance Information Criterion (DIC), a measure of model complexity and precision introduced by (Spiegelhalter et al., 2002), is utilized for model comparison. DIC value can be calculated as follows

$$DIC = \bar{D} + p_D$$
(53)

where $\bar{D}$ represents the posterior mean deviance that measures the model fitting, $p_D$ is a measure of model complexity estimating the effective number of parameters. Typically, a smaller DIC suggests a more favorable model. Models are considered competitive when the difference in DIC values is less than 5, whereas a difference exceeding 10 can decisively favor the model with the lower DIC value (Guo et al., 2020b; Yue et al., 2024).

## 3.5. Crash estimation

As shown in **Figs. 1-5**, the hybrid structure models enable the derivation of safety metrics such as the risk of crash from extreme conflicts. Specifically, regarding traffic conflicts assessed by post encroachment time (PET), a smaller PET value indicates an increased likelihood of a traffic event resulting in a crash. When PET is less than or equal to zero, a crash is considered to be occurred. Therefore, the crash risk can be calculated from the fitted hybrid GPD model of the negated PET values (Songchitruksa and Tarko, 2006; Zheng et al., 2014a). The risk of crash can be computed as

$$R_j = \Pr(\text{PET}_j < 0) = \Pr(-\text{PET}_j > 0) = 1 - G_j(0) = \begin{cases} \left(1 + \xi_j \dfrac{0 - \mu_j}{\sigma_j}\right)^{-1/\xi_j} & \xi_j \neq 0 \\ \exp\left(-\dfrac{0 - \mu_j}{\sigma_j}\right) & \xi_j = 0 \end{cases} \quad (54)$$

where $R_j$ denotes the crash risk in $j$th cycle. $G_j(\cdot)$ is the PDF of GPD in $j$th cycle.

Under the assumption that the observed samples over the given time accurately reflect the characteristics of the overall population, crashes for a longer time period $T$ (typically $T = 1$ year) can be estimated as

$$C = \frac{T}{t} \sum_{j=1} R_j \quad (55)$$

where $t$ is the duration of observation; $R_j$ is the crash risk of the $j$-th cycle during the observation period; $C$ is the estimated count of crashes over the period $T$.

## 4. Model Application

### 4.1. Data description

The BHHM models were applied for modeling rear-end traffic conflicts at signalized intersections. The data used in this study were collected from three intersections in Surrey, Canada. They are 72 Ave & 128 St, 72 Ave & 132 St, and Fraser Hwy & 168

St. Video data capturing the entrance areas of the intersections, specifically targeting rear-end conflicts were collected through cameras mounted on posts. The video data collection locations for the intersections are shown in **Fig. 6**.

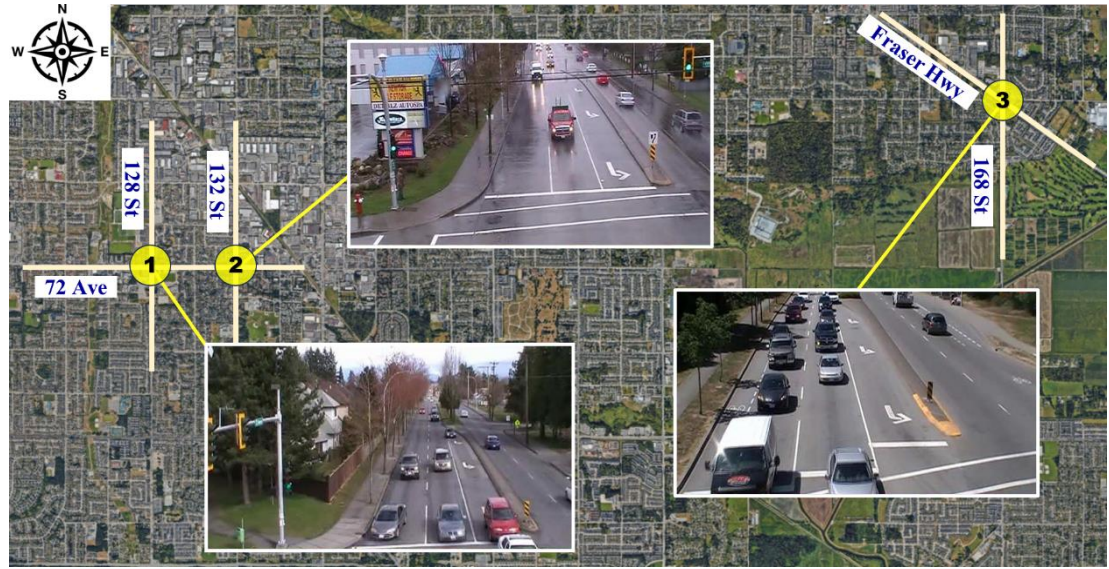

Fig. 6. Intersection locations for video collection

The video data was processed using an automated traffic conflict analysis system, employing computer vision techniques to extract rear-end conflicts at the intersections. First, camera calibration mapped the 2D image space to 3D real-world coordinates for transferring the unit of pixel into the real-world unit. Feature tracking was then performed using the Kanade-Lucas-Tomasi (KLT) algorithm to track moving vehicle features. Features at the same motion pattern are grouped into coherent objects, and the vehicle trajectories are extracted. A Kalman filter was applied to smooth the trajectory data and reduce noise. Motion prototypes were created to describe vehicle movement patterns, and trajectories were matched to these prototypes using the Longest Common Subsequence (LCSS) algorithm. Finally, conflicts were identified by analyzing the interactions between the matched trajectories. Detailed video processing procedures was outlined in [Essa and Sayed (2018)](). From the extracted vehicle trajectories, the conflict indicator PET was automatically computed. For each pair of vehicles in the through lane, the PET was measured and recorded. Furthermore, three traffic parameters at the signal cycle level, namely traffic volume (V, vehicles per lane per cycle), shock wave area (A, km·s), and platoon ratio (P), were extracted from the space-time diagram. In this study, a PET threshold of 4 seconds was used to classify vehicle interactions as traffic conflicts. Vehicle interactions with a PET greater than 4 seconds were not considered conflicts, as they are less likely to result in a collision (Zheng et al., 2019a; Fu et al., 2020; Lanzaro et al., 2023). In total, 11 hours of video footage were collected at these intersections, resulting in the extraction of 3,310 PET-related conflicts across 328 signal cycles. **Fig. 7** illustrates the histogram of PET values at the intersections of 72 Ave & 128 St, 72 Ave & 132 St, and Fraser Hwy & 168 St, respectively. The specific data collection sites and results are summarized in **Table 1**.

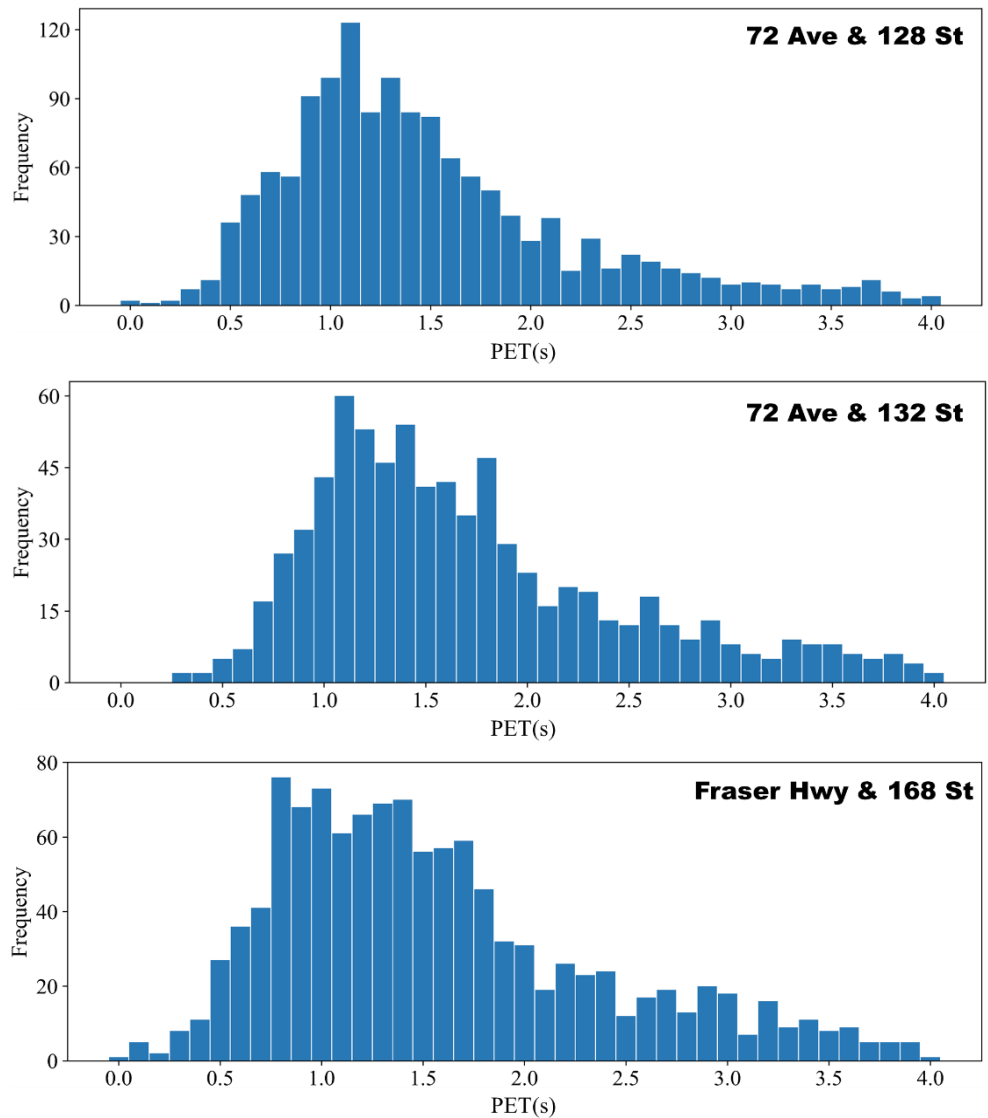

Figure. 7. Histogram for PETs in three intersections

A three-year dataset (2013–2015) of crash records from the Insurance Corporation of British Columbia (ICBC) was utilized for the three signalized intersections. According to Guo and Sayed (2020), these records contained various parameters such as location, date, time, crash type, and crash severity. This study focused in rear-end crashes occurring at the through lanes of the specified approaches. Moreover, given that rear-end crashes were primarily observed during daylight hours, only rear-end crashes during daylight hours were considered. Ultimately, a total of 43 rear-end crashes were extracted, as detailed in **Table 1**.

Table 1. Summary statistics of data.

| Site | | 72 Ave & 128 St | 72 Ave & 132 St | Fraser Hwy & 168 St |
|---|---|---|---|---|
| Video recording time | | March 28, 2012 (10:00–12:00; 14:00–16:00) | April 3, 2012 (09:00–12:00) | June 10, 2015 (09:00–13:00) |
| Video duration (h) | | 4 | 3 | 4 |
| Selected intersection approach | | Eastbound & Westbound | Westbound | Southbound |
| Number of cycles | | 133 | 79 | 116 |
| Number of conflicts | | 992 | 327 | 601 |
| PET(s) | Min | 0.12 | 0.15 | 0.11 |
| | Max | 3.82 | 3.75 | 3.74 |
| | Mean | 1.38 | 1.61 | 1.51 |
| V (vehicle count per cycle) | Min | 2 | 6 | 3 |
| | Max | 28 | 25 | 23 |
| | Mean | 15.83 | 14.51 | 15.16 |
| A (km·s) | Min | 0.00 | 0.00 | 0.00 |
| | Max | 4.45 | 2.90 | 5.66 |
| | Mean | 1.79 | 0.97 | 1.27 |
| P | Min | 0.00 | 0.14 | 0.32 |
| | Max | 1.88 | 1.81 | 1.75 |
| | Mean | 1.11 | 1.13 | 1.12 |
| Observed crashes | 2013 | 8 | 0 | 4 |
| | 2014 | 3 | 1 | 3 |
| | 2015 | 3 | 0 | 3 |
| | Mean | 4.7 | 0.3 | 3.3 |

## 4.2. Modeling results

Using the WinBUGS tool, the Markov Chain Monte Carlo (MCMC) technique was employed to sample the posterior distribution, allowing estimation of the posterior mean and standard deviation of the model parameters. Several methods can be used to monitor the convergence. Initially, two parallel chains with different initial values were tracked to ensure complete coverage of the sample space. Additionally, the Brooks-Gelman-Rubin statistic was employed, and convergence was assumed if the value of this statistic is less than 1.2. Visual inspection of MCMC trace plots of model parameters further confirmed convergence. In this study, each model's two independent Markov chains were run for 80,000 iterations, with the initial 40,000 iterations utilized for convergence monitoring and subsequently discarded as burn-in samples. The remaining iterations were used for parameter estimation.

In this study, five non-stationary BHHM models were developed and estimated. **Tables 2 - 4** present the parameter estimation results for these models. The DIC values are 6952.6, 6963.7, 7280.9, 4468.2, and 3484.2, respectively. The results indicate that the Lognormal-GPD model demonstrates the best goodness of fit with the lowest DIC among the five models. Moreover, the quality of the best fitted model is supported by the Q-Q (quantile-quantile) plots, P-P (probability-probability) plots, and kernel probability density plots as shown in **Fig. 8**.

Notably, the DIC values of the Normal-GPD model, Cauchy-GPD model, and Logistic-GPD model are close to each other but significantly higher than those of the Gamma-GPD model and Lognormal-GPD model. This suggests that the Normal, Cauchy, and Logistic distributions may not effectively capture the characteristics of general conflicts. Specifically, the normal, logistic, and Cauchy distributions, all of which are symmetric, struggle to effectively capture the skewed nature of general conflicts. Despite having heavier tails, these distributions fail to accommodate the pronounced asymmetry typically observed in general conflict data. In contrast, the gamma and lognormal distributions, which naturally handle skewness and exhibit more flexible tail behaviors, provide a better fit for the asymmetric distribution of general conflicts, allowing the hybrid structure models to achieve superior goodness of fit.

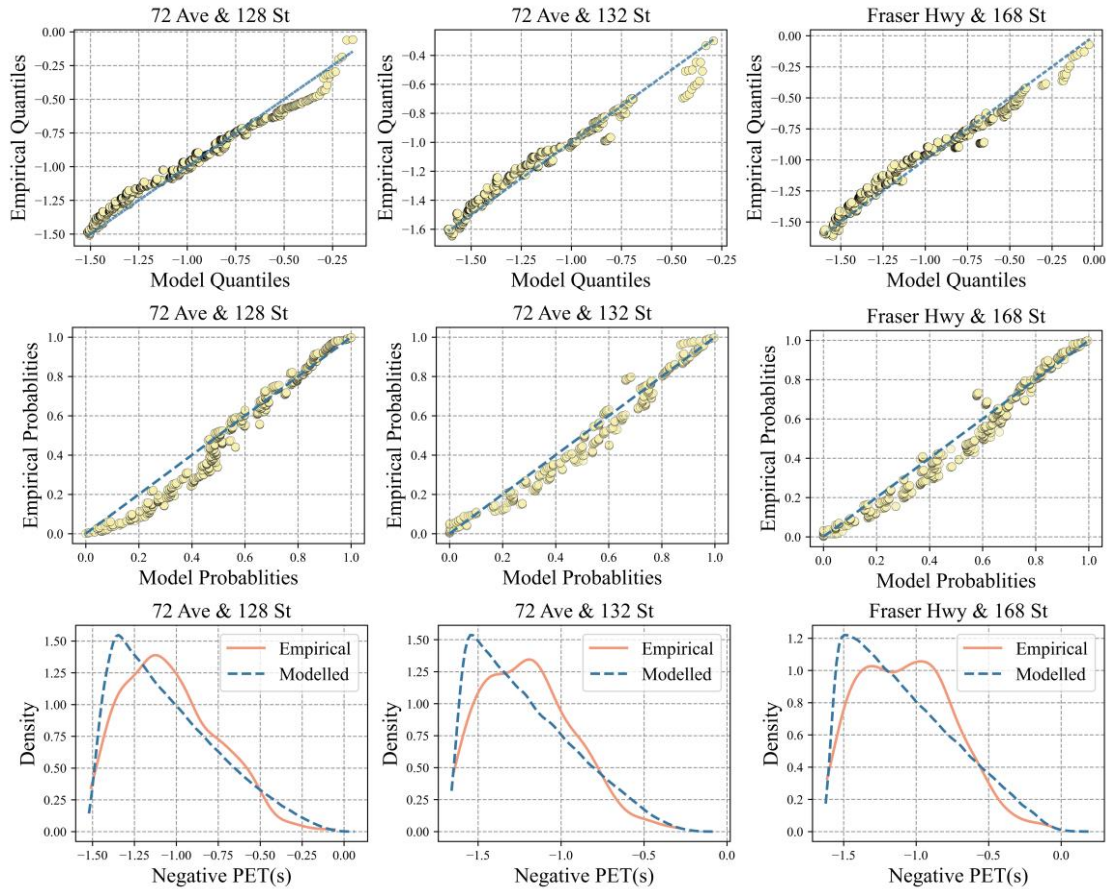

Figure 8. Q-Q plots, P-P plots, and kernel probability density plots for the best-fitted model (Lognormal-GPD model)

Moreover, the DIC difference between the Gamma-GPD model and the Lognormal-GPD model is nearly 1000. This is unexpected, given that both models are designed to handle skewed and heavy-tailed general conflict distributions. This significant discrepancy can be attributed to differences in the distribution shapes and how these distributions handle the data characteristics. The gamma distribution, with its strong skewness and relatively lighter tails, is suitable for datasets with moderate skewness and fewer extreme values. In contrast, the lognormal distribution, while also capable of modeling skewed data, features a heavier tail that is better suited for accommodating extreme values, which are modeled by GPD. This flexibility in tail behavior allows the Lognormal-GPD model to more effectively capture extreme conflict data. The Lognormal distribution's ability to handle both skewness and heavy tails likely contributes to the superior performance of the Lognormal-GPD model in capturing extreme values.

Table 2. Estimation results of the hybrid Normal-GPD model and the hybrid Cauchy-GPD model.

| Normal-GPD model | | | | | | Cauchy-GPD model | | | | | |
|---|---|---|---|---|---|---|---|---|---|---|---|
| Model parameter | | Mean | S.D. | 2.5%C.I. | 97.5%C.I. | Model parameter | | Mean | S.D. | 2.5%C.I. | 97.5%C.I. |
| $\mu$ | $\beta_{\mu 0}[1]$ | -1.662 | 0.026 | -1.706 | -1.591 | $\mu$ | $\beta_{\mu 0}[1]$ | -1.502 | 0.014 | -1.515 | -1.453 |
| | $\beta_{\mu 0}[2]$ | -1.537 | 0.030 | -1.575 | -1.474 | | $\beta_{\mu 0}[2]$ | -1.464 | 0.020 | -1.507 | -1.414 |
| | $\beta_{\mu 0}[3]$ | -1.466 | 0.019 | -1.508 | -1.434 | | $\beta_{\mu 0}[3]$ | -1.086 | 0.014 | -1.106 | -1.067 |
| | $\beta_{\mu 0}(A)$ | 0.082 | 0.010 | 0.065 | 0.099 | | $\beta_{\mu 0}(A)$ | - | - | - | - |
| $\varphi = log(\sigma)$ | $\beta_{\varphi 0}[1]$ | -0.382 | 0.056 | -0.492 | -0.272 | $\varphi = log(\sigma)$ | $\beta_{\varphi 0}[1]$ | -0.476 | 0.064 | -0.617 | -0.370 |
| | $\beta_{\varphi 0}[2]$ | -0.746 | 0.086 | -0.935 | -0.592 | | $\beta_{\varphi 0}[2]$ | -0.752 | 0.080 | -0.916 | -0.604 |
| | $\beta_{\varphi 0}[3]$ | -0.523 | 0.058 | -0.635 | -0.404 | | $\beta_{\varphi 0}[3]$ | -1.067 | 0.079 | -1.229 | -0.913 |
| | $\beta_{\varphi 0}(V)$ | - | - | - | - | | $\beta_{\varphi 0}(V)$ | 0.017 | 0.003 | 0.012 | 0.022 |
| | $\beta_{\varphi 0}(A)$ | -0.063 | 0.017 | -0.097 | -0.031 | | $\beta_{\varphi 0}(A)$ | - | - | - | - |
| | $\beta_{\varphi 0}(P)$ | - | - | - | - | | $\beta_{\varphi 0}(P)$ | -0.216 | 0.026 | -0.269 | -0.163 |
| $\xi$ | $\beta_{\xi 0}[1]$ | -0.405 | 0.020 | -0.444 | -0.365 | $\xi$ | $\beta_{\xi 0}[1]$ | -0.482 | 0.023 | -0.526 | -0.437 |
| | $\beta_{\xi 0}[2]$ | -0.353 | 0.039 | -0.423 | -0.269 | | $\beta_{\xi 0}[2]$ | -0.406 | 0.037 | -0.474 | -0.328 |
| | $\beta_{\xi 0}[3]$ | -0.403 | 0.028 | -0.458 | -0.346 | | $\beta_{\xi 0}[3]$ | -0.289 | 0.037 | -0.354 | -0.208 |
| $\kappa$ | $\beta_{\kappa 0}[1]$ | -1.117 | 0.119 | -1.348 | -0.879 | $x_0$ | $\beta_{x00}[1]$ | -1.291 | 0.034 | -1.358 | -1.223 |
| | $\beta_{\kappa 0}[2]$ | -1.429 | 0.113 | -1.642 | -1.205 | | $\beta_{x00}[2]$ | -1.575 | 0.038 | -1.648 | -1.499 |
| | $\beta_{\kappa 0}[3]$ | -1.309 | 0.111 | -1.522 | -1.088 | | $\beta_{x00}[3]$ | -1.488 | 0.028 | -1.542 | -1.434 |
| | $\beta_{\kappa 0}(V)$ | 0.018 | 0.005 | 0.008 | 0.027 | | $\beta_{\kappa 0}(V)$ | - | - | - | - |
| | $\beta_{\kappa 0}(P)$ | -0.358 | 0.071 | -0.500 | -0.225 | | $\beta_{\kappa 0}(P)$ | - | - | - | - |
| $\hat{\lambda} = log(\lambda)$ | $\beta_{\lambda 0}[1]$ | 0.063 | 0.036 | -0.007 | 0.133 | $\hat{\gamma} = log(\gamma)$ | $\beta_{\gamma 0}[1]$ | -0.559 | 0.038 | -0.637 | -0.484 |
| | $\beta_{\lambda 0}[2]$ | -0.021 | 0.039 | -0.098 | 0.057 | | $\beta_{\gamma 0}[2]$ | -0.587 | 0.044 | -0.675 | -0.504 |
| | $\beta_{\lambda 0}[3]$ | 0.012 | 0.031 | -0.048 | 0.074 | | $\beta_{\gamma 0}[3]$ | -0.627 | 0.030 | -0.684 | -0.569 |
| DIC | | 6952.6 | | | | DIC | | 6963.7 | | | |

Notes: '-' indicates that the covariate is not included, 'V' stands for traffic volume, 'A' denotes the shock wave area, 'P' denotes the platoon ratio.

Table 3. Estimation results of the hybrid Logistic-GPD model and the hybrid Gamma-GPD model.

| **Logistic-GPD model** | | | | | | **Gamma-GPD model** | | | | | |
|---|---|---|---|---|---|---|---|---|---|---|---|
| **Model parameter** | | **Mean** | **S.D.** | **2.5%C.I.** | **97.5%C.I.** | **Model parameter** | | **Mean** | **S.D.** | **2.5%C.I.** | **97.5%C.I.** |
| $\mu$ | $\beta_{\mu 0}[1]$ | -1.198 | 0.011 | -1.211 | -1.167 | $\mu$ | $\beta_{\mu 0}[1]$ | -1.287 | 0.001 | -1.290 | -1.285 |
| | $\beta_{\mu 0}[2]$ | -1.269 | 0.008 | -1.284 | -1.246 | | $\beta_{\mu 0}[2]$ | -1.311 | 0.003 | -1.316 | -1.307 |
| | $\beta_{\mu 0}[3]$ | -1.081 | 0.013 | -1.105 | -1.067 | | $\beta_{\mu 0}[3]$ | -1.114 | 0.003 | -1.125 | -1.111 |
| | $\beta_{\mu 0}(A)$ | - | - | - | - | | $\beta_{\mu 0}(A)$ | 0.022 | 0.001 | 0.021 | 0.023 |
| $\varphi=log(\sigma)$ | $\beta_{\varphi 0}[1]$ | -1.215 | 0.068 | -1.378 | -1.111 | $\varphi=log(\sigma)$ | $\beta_{\varphi 0}[1]$ | -0.792 | 0.074 | -0.942 | -0.643 |
| | $\beta_{\varphi 0}[2]$ | -1.347 | 0.102 | -1.548 | -1.148 | | $\beta_{\varphi 0}[2]$ | -1.024 | 0.097 | -1.214 | -0.838 |
| | $\beta_{\varphi 0}[3]$ | -1.260 | 0.092 | -1.432 | -1.075 | | $\beta_{\varphi 0}[3]$ | -0.975 | 0.086 | -1.142 | -0.800 |
| | $\beta_{\varphi 0}(V)$ | 0.014 | 0.004 | 0.008 | 0.021 | | $\beta_{\varphi 0}(V)$ | - | - | - | - |
| | $\beta_{\varphi 0}(P)$ | - | - | - | - | | $\beta_{\varphi 0}(P)$ | -0.070 | 0.052 | -0.174 | -0.016 |
| $\xi$ | $\beta_{\xi 0}[1]$ | -0.288 | 0.023 | -0.323 | -0.237 | $\xi$ | $\beta_{\xi 0}[1]$ | -0.330 | 0.024 | -0.372 | -0.279 |
| | $\beta_{\xi 0}[2]$ | -0.274 | 0.055 | -0.370 | -0.153 | | $\beta_{\xi 0}[2]$ | -0.272 | 0.048 | -0.359 | -0.166 |
| | $\beta_{\xi 0}[3]$ | -0.301 | 0.040 | -0.376 | -0.214 | | $\beta_{\xi 0}[3]$ | -0.279 | 0.040 | -0.350 | -0.193 |
| $\vartheta$ | $\beta_{\vartheta 0}[1]$ | -1.403 | 0.019 | -1.440 | -1.363 | $\hat{p}=log(p)$ | $\beta_{p 0}[1]$ | 2.967 | 0.029 | 2.909 | 3.024 |
| | $\beta_{\vartheta 0}[2]$ | -1.604 | 0.027 | -1.658 | -1.550 | | $\beta_{p 0}[2]$ | 2.886 | 0.035 | 2.819 | 2.956 |
| | $\beta_{\vartheta 0}[3]$ | -1.483 | 0.022 | -1.528 | -1.439 | | $\beta_{p 0}[3]$ | 2.698 | 0.029 | 2.638 | 2.751 |
| $\hat{g}=log(g)$ | $\beta_{g 0}[1]$ | -0.972 | 0.113 | -1.213 | -0.755 | $\hat{q}=log(q)$ | $\beta_{q 0}[1]$ | 2.237 | 0.032 | 2.172 | 2.300 |
| | $\beta_{g 0}[2]$ | -0.889 | 0.108 | -1.105 | -0.684 | | $\beta_{q 0}[2]$ | 2.108 | 0.039 | 2.036 | 2.184 |
| | $\beta_{g 0}[3]$ | -0.860 | 0.105 | -1.067 | -0.663 | | $\beta_{q 0}[3]$ | 1.991 | 0.032 | 1.925 | 2.051 |
| | $\beta_{g 0}(V)$ | 0.015 | 0.006 | 0.003 | 0.028 | | $\beta_{q 0}(V)$ | - | - | - | - |
| DIC | | 7280.9 | | | | DIC | | 4468.2 | | | |

Notes: '-' indicates that the covariate is not included, 'V' stands for traffic volume, 'A' denotes the shock wave area, 'P' denotes the platoon ratio.

Table 4. Estimation results of the hybrid Lognormal-GPD model.

| Model parameter | | Mean | S.D. | 2.5%C.I. | 97.5%C.I. |
|---|---|---|---|---|---|
| $\mu$ | $\beta_{\mu 0}[1]$ | -1.520 | 0.017 | -1.541 | -1.461 |
| | $\beta_{\mu 0}[2]$ | -1.677 | 0.043 | -1.772 | -1.636 |
| | $\beta_{\mu 0}[3]$ | -1.626 | 0.037 | -1.718 | -1.580 |
| | $\beta_{\mu 0}(A)$ | 0.036 | 0.007 | 0.022 | 0.048 |
| $\varphi = log(\sigma)$ | $\beta_{\varphi 0}[1]$ | -0.522 | 0.048 | -0.619 | -0.430 |
| | $\beta_{\varphi 0}[2]$ | -0.470 | 0.080 | -0.610 | -0.299 |
| | $\beta_{\varphi 0}[3]$ | -0.281 | 0.063 | -0.397 | -0.150 |
| | $\beta_{\varphi 0}(A)$ | -0.024 | 0.014 | -0.049 | 0.004 |
| $\xi$ | $\beta_{\xi 0}[1]$ | -0.393 | 0.019 | -0.429 | -0.355 |
| | $\beta_{\xi 0}[2]$ | -0.439 | 0.033 | -0.502 | -0.374 |
| | $\beta_{\xi 0}[3]$ | -0.475 | 0.026 | -0.525 | -0.424 |
| $v$ | $\beta_{v 0}[1]$ | 0.765 | 0.026 | 0.712 | 0.814 |
| | $\beta_{v 0}[2]$ | 0.879 | 0.031 | 0.824 | 0.949 |
| | $\beta_{v 0}[3]$ | 0.851 | 0.029 | 0.797 | 0.913 |
| $\hat{w} = log(w)$ | $\beta_{w 0}[1]$ | -1.421 | 0.026 | -1.470 | -1.368 |
| | $\beta_{w 0}[2]$ | -1.483 | 0.039 | -1.565 | -1.413 |
| | $\beta_{w 0}[3]$ | -1.477 | 0.035 | -1.554 | -1.416 |
| DIC | | 3490.2 | | | |

Notes: 'A' denotes the shock wave area.

## 5. Discussions

### 5.1. Threshold determination in peak over threshold method

In the EVT peak over threshold method, it is crucial to select an appropriate threshold to distinguish extremes and general events, where such threshold have great influence on the accuracy and precision of the estimation. The choice of the threshold is subject to a bias-variance trade-off (Northrop et al.,2017). If the threshold is set too high, very few samples can be used for the parameters estimation, resulting in high variance in the estimator. Conversely, if the threshold is too low, the GPD approximation of the tail will exhibit a large bias. According to Pickands' theorem, convergence occurs as the threshold approaches the right endpoint of the distribution (Bercher and Vignat, 2008). As such, it is essential to choose the threshold in an objective way (Scarrott and MacDonald, 2012; Yue et al., 2024).

As shown in **Tables 2-4**, the threshold parameter $\mu$ is estimated in an objective way from the BHHM models. The proposed threshold estimation approach overcomes the drawback of determining threshold in a subjective and arbitrary way. Besides, most previous studies assumed that the threshold in the peak over threshold method is constant for a specific site (Guo et al., 2020a; Zheng et al., 2019b). However, this assumption might be inaccurate since in different traffic environments, the distribution of traffic conflicts and the traffic status may vary across traffic status with varied traffic flow, even at the same site. As such, the scenario of treating the threshold as a signal parameter was compared with the scenario of treating the threshold as a link function of covariates. The results are shown in **Figs. 10-12**. The results showed that the estimated threshold in the non-stationary Lognormal-GPD model varies across signal cycles, while the threshold in the stationary Lognormal-GPD model keeps consistent. The result indicates that the non-stationary model is able to capture the characteristics of threshold variation across signal cycles. Moreover, the threshold used in the non-stationary model is generally smaller than the counterpart threshold used in the stationary model. As shown in **Figs. 10-12**, for the stationary threshold, the extreme conflicts counts are 702, 330, and 579 for 72 Ave & 128 St, 72 Ave & 132 St, and Fraser Hwy & 168 St, respectively. For the non-stationary threshold, the exceedances are 762, 407, and 657 at the same sites, respectively.

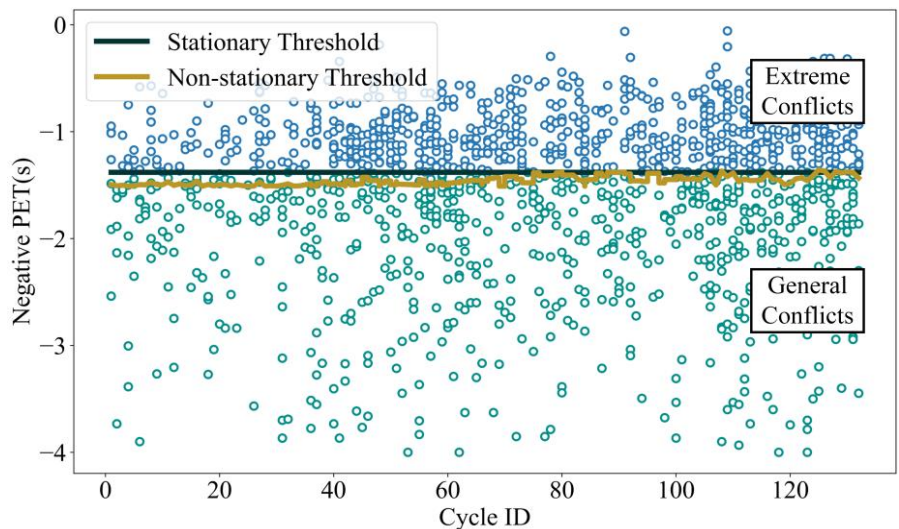

Figure 10. Difference between stationary and non-stationary threshold (72 Ave & 128 St)

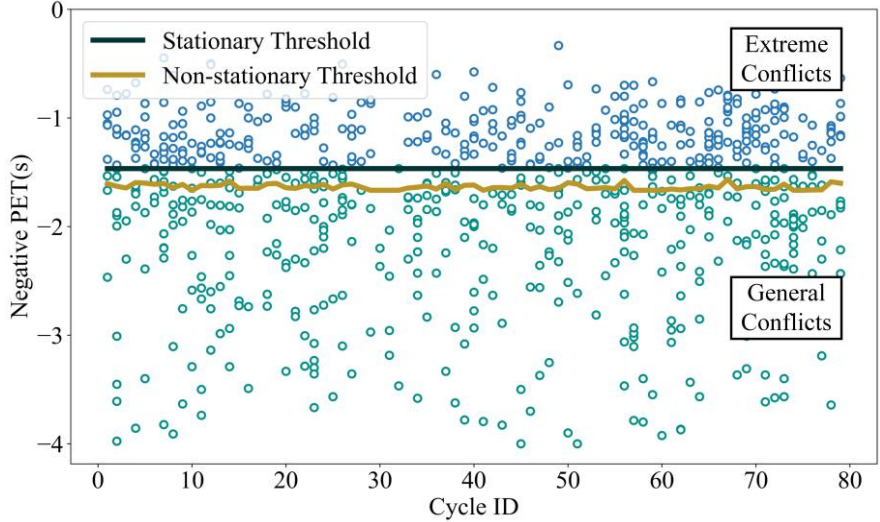
Figure 11. Difference between stationary and non-stationary threshold (72 Ave & 132 St)

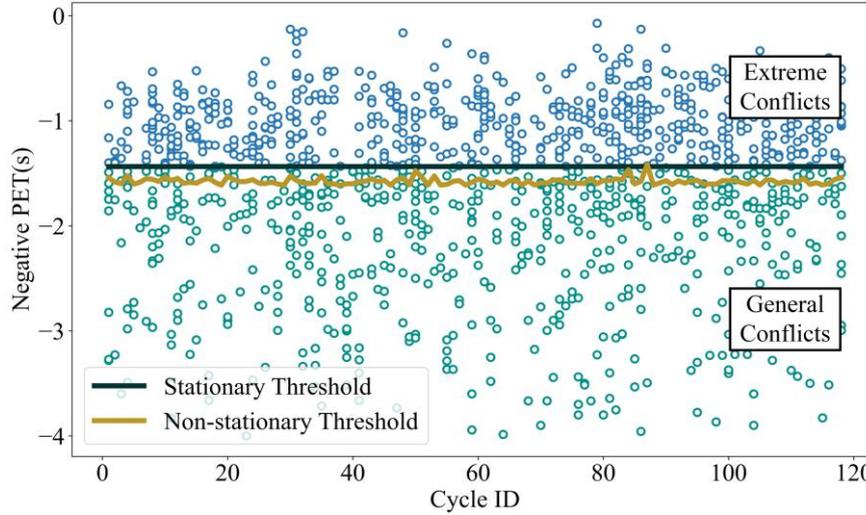
Figure 12. Difference between stationary and non-stationary threshold (Fraser Hwy & 168 St)

## 5.2. Comparison with graphical diagnostics and quantile regression approaches

To verify the effectiveness of the proposed non-stationary BHHM models for determining the threshold $\mu$, the models were compared against the traditional graphical diagnostics method and quantile regression approach. The performance of the three methods was evaluated by comparing the annual number of crashes estimated from the traffic conflict data with the actual observed crash data. The 95% C.I. were also used for comparison. According to Songchitruksa and Tarko (2006), the number of crashes is assumed following a Poisson distribution. Hence, the 95% C.I. for the observed crash mean can be calculated as $\left\{\lambda: \frac{1}{2n}\chi^2_{2y_0, 0.975} \leq \lambda \leq \frac{1}{2n}\chi^2_{2(y_0+1), 0.025}\right\}$, where $n = 3$ refers the number of years for crash data collection, and $y_0$ represents the total number of crashes in the 3-year period for specific site.

**5.2.1. Graphical diagnostics approach**

The traditional graphical diagnostics approach determines the threshold in a relatively subjective manner. Specifically, the mean residual life plot and threshold stability plot are used for threshold selection. A threshold is identified when the mean residual life plot exhibits approximate linearity and the threshold stability plot shows near constancy (Reiss et al., 1997; Scarrott and MacDonald, 2012). As an illustrative example, the intersection of 72 Ave & 128 St was analyzed, as shown in **Fig. 13**. The figure indicates that the mean residual life plot exhibits near-linearity over the ranges [-4, -0.5], while both the modified scale and shape parameters remain relatively constant within the range [-1.1, -0.9]. The overlapping of the above two ranges, i.e. [-1.1, -0.9], represents the range of potential thresholds. Typically, the maximum value within this range is selected as the final threshold. Therefore, -0.9 was chosen as the threshold for use at the intersection of 72 Ave & 128 St. Similarly, thresholds of -1.2 and -0.9 were selected for use at the intersections of 72 Ave & 132 St and Fraser Hwy & 168 St, respectively. The corresponding mean residual life and threshold stability plots are presented in **Fig. 14** and **Fig. 15**. The Bayesian hierarchical GPD model based on the calculated threshold were developed. **Table 5** presents the parameter estimation results based on the graphical diagnostics approach.

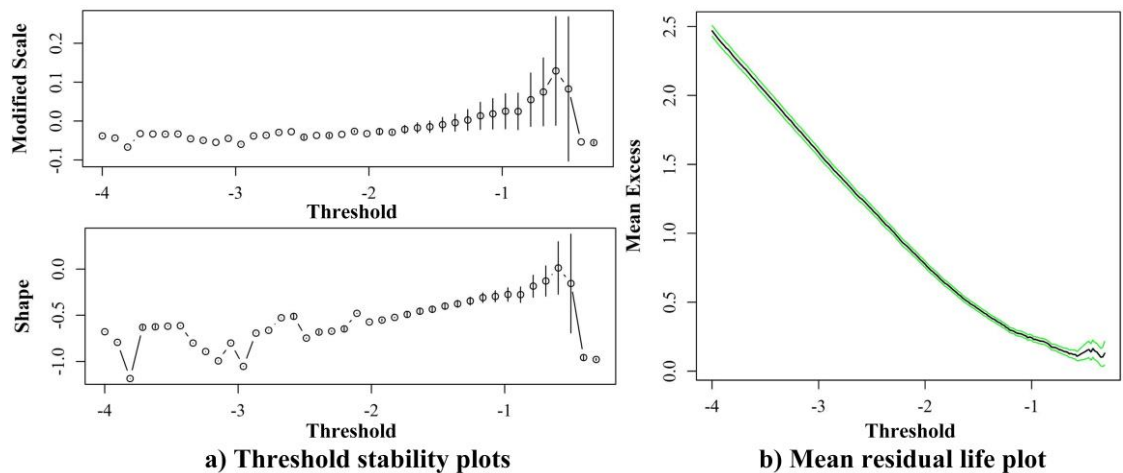

Figure 13. Threshold selection diagnostic plots (72 Ave & 128 St)

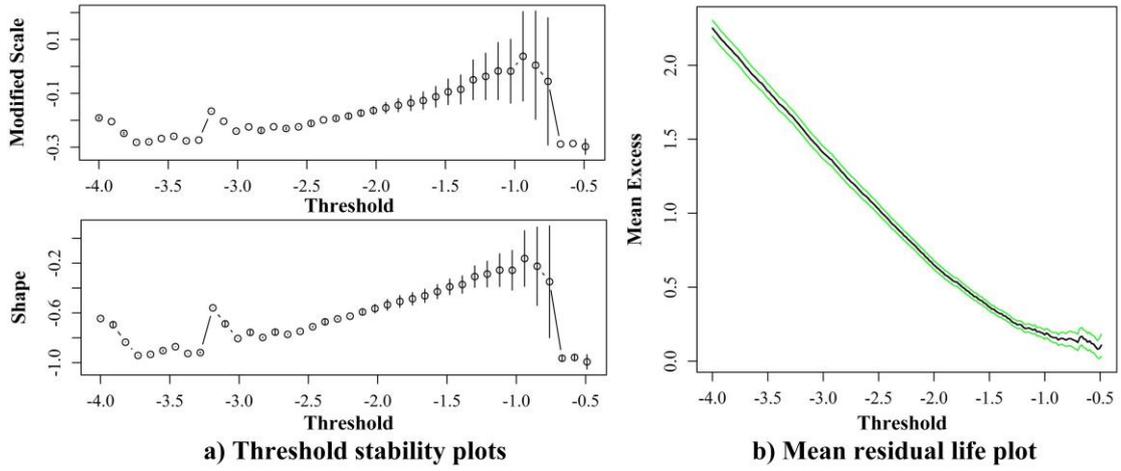

Figure 14. Threshold selection diagnostic plots (72 Ave & 132 St)

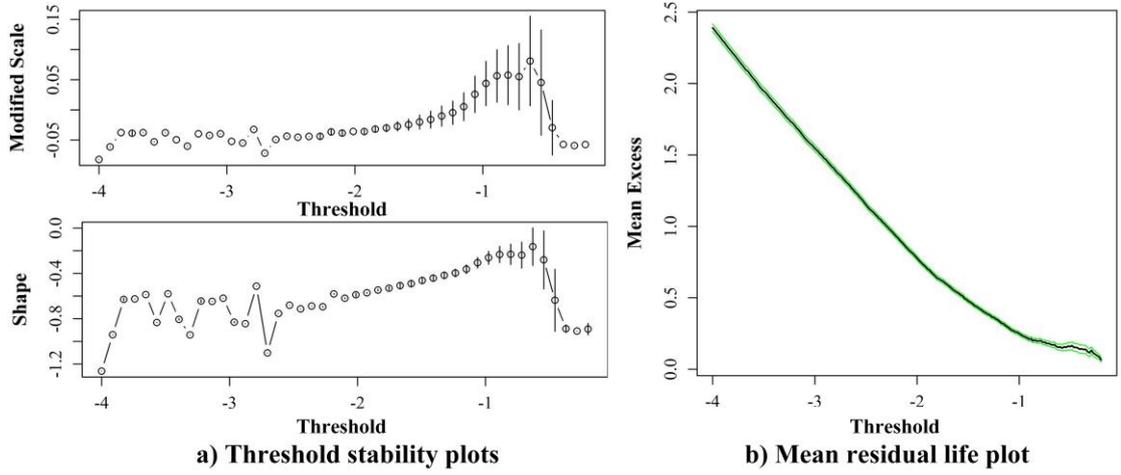

Figure 15. Threshold selection diagnostic plots (Fraser Hwy & 168 St)

Table 5. Model estimation results for GPD models based on graphical diagnostics approach

| Model parameter | | Mean | S.D. | 2.5% C.I. | 97.5% C.I. |
|---|---|---|---|---|---|
| $\mu$ | $\beta_{\mu 0}[1]$ | -0.9 | | | |
| | $\beta_{\mu 0}[2]$ | -1.2 | | | |
| | $\beta_{\mu 0}[3]$ | -0.9 | | | |
| $\varphi = log(\sigma)$ | $\beta_{\varphi 0}[1]$ | -1.188** | 0.157 | -1.528 | -0.900 |
| | $\beta_{\varphi 0}[2]$ | -1.131** | 0.143 | -1.403 | -0.849 |
| | $\beta_{\varphi 0}[3]$ | -1.378** | 0.148 | -1.680 | -1.098 |
| | $\beta_{\varphi 0}(V)$ | 0.036** | 0.008 | 0.020 | 0.052 |
| | $\beta_{\varphi 0}(A)$ | -0.094** | 0.031 | -0.147 | -0.030 |
| | $\beta_{\varphi 0}(P)$ | -0.435** | 0.099 | -0.617 | -0.242 |
| $\xi$ | $\beta_{\xi 0}[1]$ | -0.357** | 0.058 | -0.475 | -0.237 |
| | $\beta_{\xi 0}[2]$ | -0.322** | 0.067 | -0.443 | -0.182 |
| | $\beta_{\xi 0}[3]$ | -0.187** | 0.064 | -0.301 | -0.050 |

Notes: '**' signify that the covariate is significant at the 0.05 levels, 'V' stands for traffic volume, 'A' denotes the shock wave area, 'P' denotes the platoon ratio.

### 5.2.2. Quantile regression approach

Quantile regression is a recently introduced technique for identifying non-stationary thresholds within the peak over threshold method (Zheng et al., 2019b; Fu and Sayed, 2023). This method estimates non-stationary thresholds by modeling the conditional quantiles of the dependent variable as linear functions of explanatory covariates. Specifically, an $\alpha$-quantile regression models the conditional $\alpha$-quantile of the dependent variable $y_\alpha$ as a function of a function of covariates $x_\alpha$, expressed as

$$Q_{(y|x)}(\alpha) = \mathbf{x}_\alpha \mathbf{\Theta}_\alpha \tag{56}$$

where $Q_{(y|x)}(\alpha)$ represents the $\alpha$-th conditional quantile of the dependent variable $y_\alpha$, $\mathbf{x}_\alpha$ denotes the vector of covariates corresponding to the $\alpha$-th quantile, and $\mathbf{\Theta}_\alpha$ is the vector of regression coefficients for the $\alpha$ th quantile, $\alpha \in (0,1)$.

The parameters of the model are estimated by minimizing the weighted absolute deviations, formulated as

$$q_\alpha = \operatorname{argmin}\left[\sum_{\alpha: q_\alpha \geq \mathbf{x}_\alpha \mathbf{\Theta}_\alpha} \alpha |q_\alpha - \mathbf{x}_\alpha \mathbf{\Theta}_\alpha| + \sum_{\alpha: q_\alpha < \mathbf{x}_\alpha \mathbf{\Theta}_\alpha} (1-\alpha)|q_\alpha - \mathbf{x}_\alpha \mathbf{\Theta}_\alpha|\right] \tag{57}$$

where $q_\alpha$ is the $\alpha$-th quantile. For a given $\alpha$, a quantile regression model can be developed.

In this study, a threshold stability plot is employed to select an appropriate upper quantile. Non-stationary thresholds for the peak over threshold approach are determined through quantile regression, with quantiles ranging from 80% to 95% at 2.5% intervals. These thresholds are modeled as functions of three covariates: traffic volume (V), shock wave area (A), and platoon ratio (P). The parameters estimation results indicate that traffic volume (V) and platoon ratio (P) is significant across the entire range of quantiles, while the shock wave area (A) covariate is not significant. **Table 6** provides the quantile regression results incorporating only the significant covariates. The findings reveal that traffic volume (V) consistently exhibited positive regression coefficients across all quantiles, while platoon ratio (P) displayed negative coefficients.

Table 6. Results of quantile regressions

| Quantile | Parameter | Coefficient | S.E. | t | p>|t| | 95%C.I. | |
|---|---|---|---|---|---|---|---|
| 80% | Intercept | -1.340** | 0.069 | -19.484 | <0.001 | -1.475 | -1.205 |
|  | V | 0.030** | 0.003 | 8.574 | <0.001 | 0.023 | 0.037 |
|  | P | -0.173** | 0.047 | -3.654 | <0.001 | -0.266 | -0.080 |
| 82.5% | Intercept | -1.327** | 0.068 | -19.471 | <0.001 | -1.461 | -1.193 |
|  | V | 0.030** | 0.003 | 8.860 | <0.001 | 0.024 | 0.037 |
|  | P | -0.163* | 0.047 | -3.463 | 0.001 | -0.255 | -0.070 |
| **85%** | **Intercept** | **-1.313**** | 0.068 | -19.282 | <0.001 | -1.447 | -1.179 |
|  | **V** | **0.031**** | 0.003 | 9.039 | <0.001 | 0.024 | 0.038 |
|  | **P** | **-0.151*** | 0.047 | -3.229 | 0.001 | -0.244 | -0.059 |
| 87.5% | Intercept | -1.292** | 0.068 | -19.010 | <0.001 | -1.425 | -1.158 |
|  | V | 0.032** | 0.003 | 9.240 | <0.001 | 0.025 | 0.038 |
|  | P | -0.146* | 0.047 | -3.118 | 0.002 | -0.238 | -0.054 |
| 90% | Intercept | -1.266** | 0.068 | -18.646 | <0.001 | 1.400 | -1.132 |
|  | V | 0.033** | 0.003 | 9.507 | <0.001 | 0.026 | 0.039 |
|  | P | -0.146* | 0.047 | -3.111 | 0.002 | -0.238 | -0.054 |
| 92.5% | Intercept | -1.244** | 0.068 | -18.333 | <0.001 | -1.377 | -1.110 |
|  | V | 0.034** | 0.003 | 10.006 | <0.001 | 0.028 | 0.041 |
|  | P | -0.148* | 0.047 | -3.173 | 0.002 | -0.240 | -0.056 |
| 95% | Intercept | -1.238** | 0.068 | -18.316 | <0.001 | -1.371 | -1.105 |
|  | V | 0.037** | 0.003 | 10.948 | <0.001 | 0.031 | 0.044 |
|  | P | -0.147* | 0.047 | -3.167 | 0.002 | -0.239 | -0.056 |

'**' signifies that the covariate is significant at the 0.001 levels, and '*' signifies that the covariate is significant at the 0.005 levels, 'V' stands for traffic volume, 'P' denotes the platoon ratio.

Based on the estimated thresholds derived from quantile regression, the scale and shape parameters of the corresponding GPD were computed for each quantile. **Fig. 16** illustrates the variation of these parameters across quantiles. The results demonstrate stable ranges for both the shape and scale parameters at the three intersections: 80%~85% for 72 Ave & 128 St, 85%~90% for 72 Ave & 132 St, and 85%~87.5% for Fraser Hwy & 168 St. Consequently, the 85% quantile regression is identified as the most suitable threshold across all three intersections. A Bayesian hierarchical GPD model based on the threshold calculated using the 85% quantile regression was developed. **Table 7** presents the non-stationary model parameters estimation results based on the quantile regression approach, the results show that the shock wave area and platoon ratio is negatively related to the scale parameter.

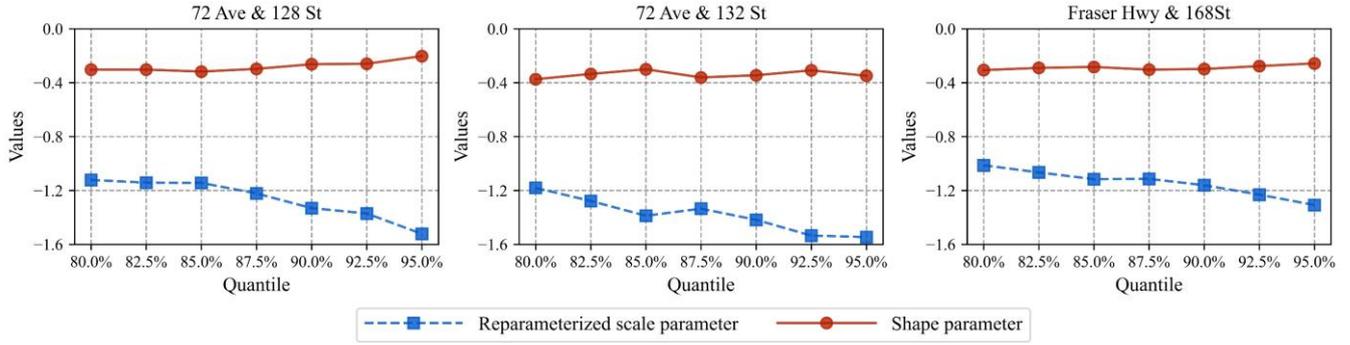

Figure 16. Threshold stability diagram.

Table 7. Parameters estimation results based on the quantile regression approach

| Model parameter | | Mean | S.D. | 2.5% C.I. | 97.5% C.I. |
|---|---|---|---|---|---|
| $\varphi=log(\sigma)$ | $\beta_{\varphi0}[1]$ | -0.709** | 0.145 | -1.025 | -0.463 |
| | $\beta_{\varphi0}[2]$ | -1.045** | 0.159 | -1.368 | -0.737 |
| | $\beta_{\varphi0}[3]$ | -0.758** | 0.123 | -1.012 | -0.536 |
| | $\beta_{\varphi0}(A)$ | -0.103** | 0.023 | -0.145 | -0.057 |
| | $\beta_{\varphi0}(P)$ | -0.206** | 0.079 | -0.345 | -0.028 |
| $\xi$ | $\beta_{\xi0}[1]$ | -0.375** | 0.049 | -0.465 | -0.276 |
| | $\beta_{\xi0}[2]$ | -0.329** | 0.092 | -0.500 | -0.131 |
| | $\beta_{\xi0}[3]$ | -0.299** | 0.046 | -0.380 | -0.201 |

Notes: '**' signify that the covariate is significant at the 0.05 levels. 'A' denotes the shock wave area. 'P' denotes the platoon ratio.

### 5.2.3. Comparison of three methods for determining threshold

**Fig. 17** illustrates the goodness-of-fit among the Lognormal-GPD model, the graphical diagnostics approach, and the quantile regression approach for the negative PET indicator using probability density plots. **Table 8** summarizes the differences in the areas under the empirical and modeled curves for these three approaches. The results indicate that both the Lognormal-GPD model and the quantile regression approach outperform the graphical diagnostics approach, demonstrating their superior accuracy in capturing the characteristics of negative PET distributions. While the quantile regression approach achieves a slightly better fit than the Lognormal-GPD model, but the difference is minimal. This slight advantage may arise because the quantile regression approach directly models the relationship between the threshold parameter and covariates through a regression framework, allowing for flexible adaptation to varying conditions. However, this method does not fully integrate this relationship into the EVT framework, which emphasizes the asymptotic properties of extreme values. As a result, this limitation might lead to biases in the prediction of extreme events.

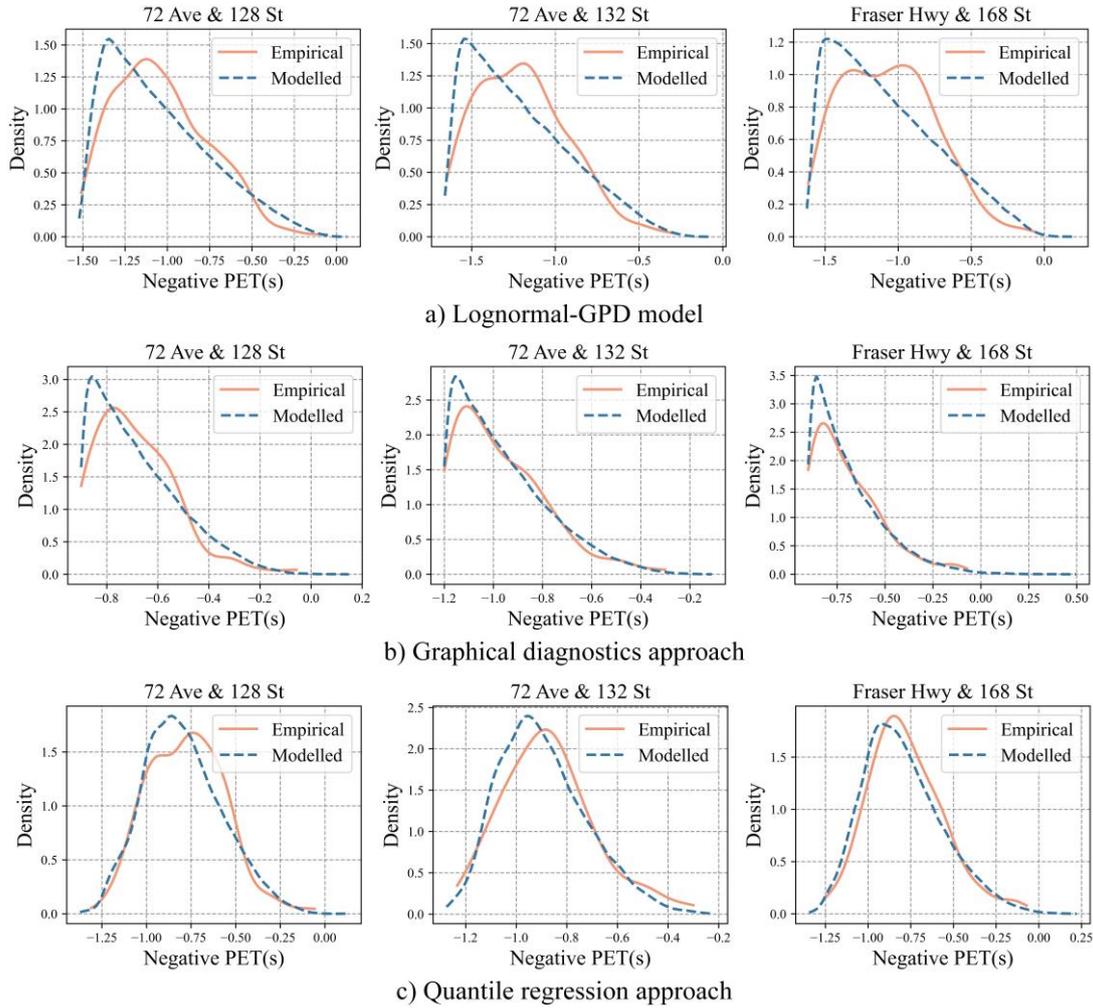

Figure 17. Kernel probability density plots

Table 8. Difference between the area under the empirical and modelled curves

| Site | 72Ave & 128St | 72Ave & 132St | Fraser Hwy & 168St |
|---|---|---|---|
| Lognormal-GPD | 0.0175 | 0.0251 | 0.0202 |
| Graphical diagnostics | 0.0422 | 0.0522 | 0.0701 |
| Quantile regression | 0.0061 | 0.0245 | 0.0137 |

**Table 9**, along with **Figs. 18-20**, presents the crash estimation results for the BHHM models, graphical diagnostic approach, and quantile regression approach. The results show that crash estimates based on the BHHM models are generally closer to the observed crashes than those derived from the graphical diagnostic and quantile regression approaches. And the 95% confidence intervals for the BHHM models are narrower compared to the other two methods, demonstrating the superior performance of the proposed BHHM model approach in crash estimation. In this study, the BHHM models incorporate both traffic conflict data below and above the threshold compared to graphical diagnostic approach and quantile regression approach. The findings show that incorporating all traffic conflict samples leads to more reliable estimates of GPD model parameters, evidenced by narrower confidence intervals (Yue et al., 2024; Hussain et al., 2022). Consequently, the estimation precision is improved.

Table 9. Comparison of estimated crashes with different threshold selection methods

| Site | 72Ave & 128St | | | 72Ave & 132St | | | Fraser Hwy & 168St | | |
|---|---|---|---|---|---|---|---|---|---|
| | Mean | 95% C.I. | | Mean | 95% C.I. | | Mean | 95% C.I. | |
| Observed | 4.7 | 2.6 | 7.8 | 0.3 | 0.0 | 1.9 | 3.3 | 1.6 | 6.1 |
| Normal-GPD | 6.0 | 0.0 | 165.7 | 0.0 | 0.0 | 2.4 | 3.7 | 0.0 | 279.0 |
| Cauchy-GPD | 6.4 | 0.6 | 24.0 | 1.4 | 0.0 | 9.0 | 110.6 | 15.1 | 401.5 |
| Logistic-GPD | 37.6 | 2.0 | 116.7 | 0.8 | 0.0 | 7.1 | 56.0 | 0.4 | 257.4 |
| Gamma-GPD | 13.2 | 0.1 | 107.7 | 2.7 | 0.0 | 27.1 | 130.5 | 3.1 | 578.4 |
| Lognormal-GPD | 5.3 | 0.0 | 51.9 | 0.0 | 0.0 | 2.1 | 3.5 | 0.0 | 37.9 |
| Graphical diagnostic | 23.8 | 0.5 | 390.1 | 0.2 | 0.0 | 257.2 | 357.2 | 24.6 | 1267.8 |
| Quantile regression | 44.1 | 9.7 | 173.9 | 6.5 | 0.0 | 67.3 | 198.8 | 47.3 | 647.9 |

In addition, the BHHM model approach demonstrates better crash estimation accuracy at the intersection of 72 Ave & 132 St, where nearly all estimated crashes fall within the 95% confidence intervals of the observed crashes. However, at the intersections of 72 Ave & 128 St and Fraser Hwy & 168 St, the BHHM model exhibits greater variability in prediction accuracy, consistently overestimating the observed crashes. A detailed examination of the crash reports reveals that only one crash occurred at 72 Ave & 132 St over three years, compared to 14 and 10 crashes at 72 Ave & 128 St and Fraser Hwy & 168 St, respectively. These results suggest that the BHHM framework is more reliable in estimating crashes at lower-risk intersections, such as 72 Ave & 132 St, but tends to overestimate crash risks at higher-risk intersections.

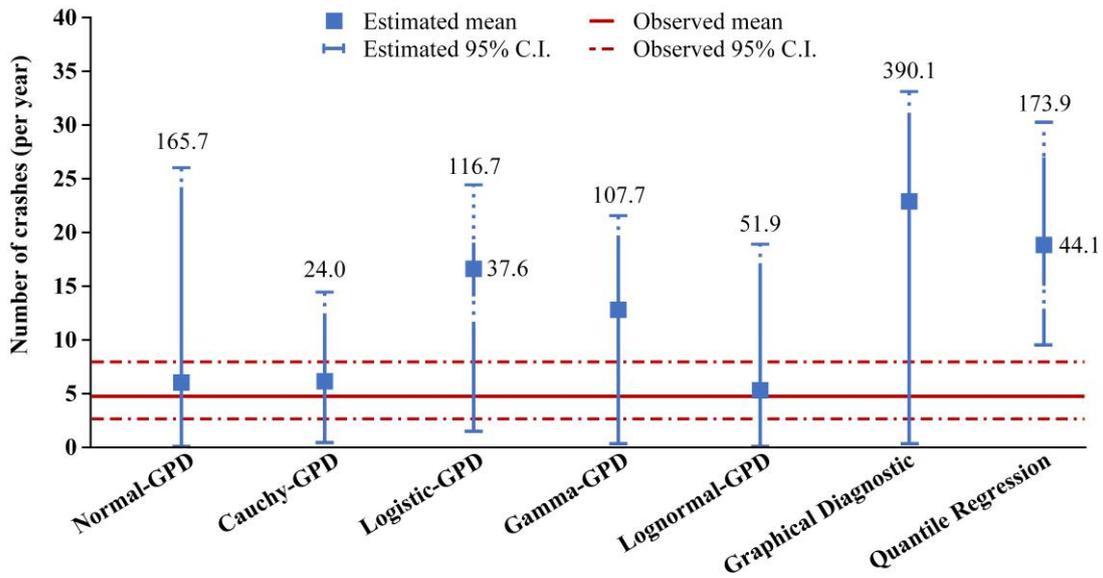

Figure 18. Comparison of estimated crashes among five hybrid GPD models, graphical diagnostic, and quantile regression approach at 72 Ave & 128 St

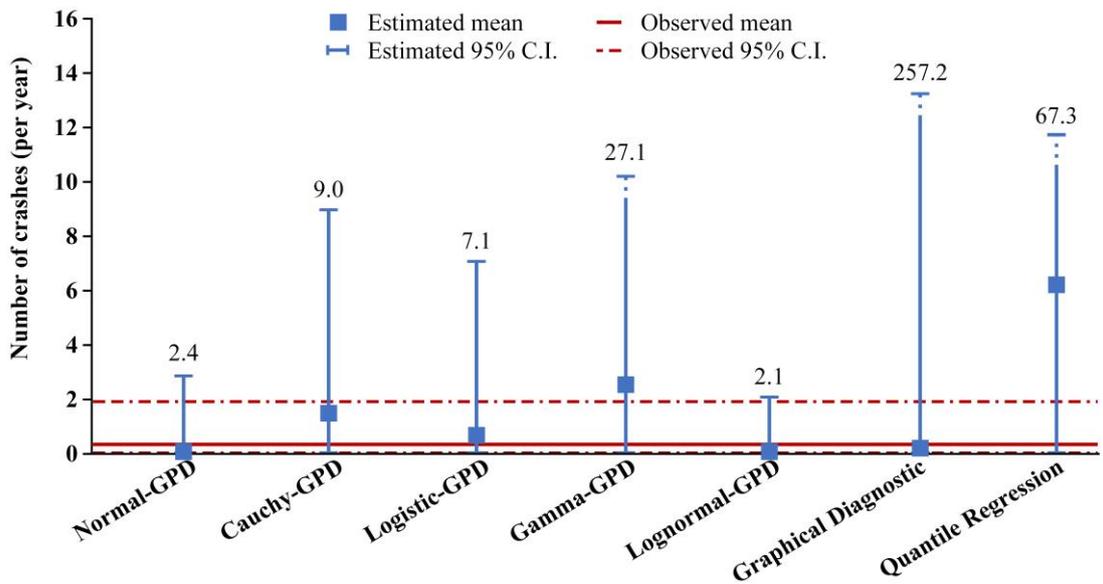

Figure 19. Comparison of estimated crashes among five hybrid GPD models, graphical diagnostic, and quantile regression approach at 72 Ave & 132 St

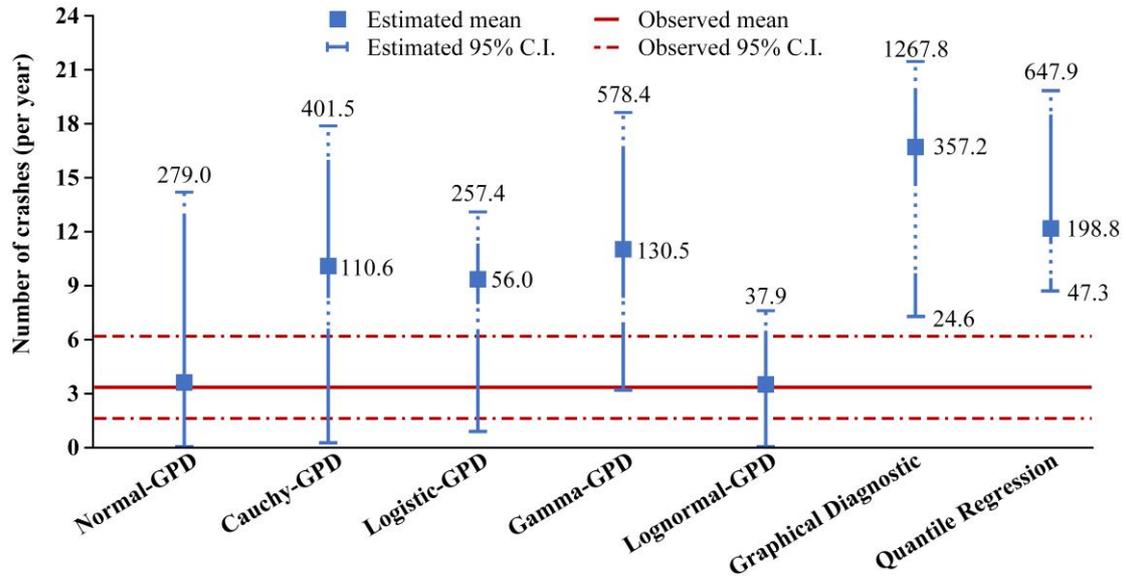

Figure 20. Comparison of estimated crashes among five hybrid GPD models, graphical diagnostic, and quantile regression approach at Fraser Highway & 168 St

## 6. Conclusions

This study develops a non-stationary BHHM framework for threshold estimation in peak over threshold approach. The BHHM framework consists of two components, a hybrid structure model and a Bayesian hierarchical structure. The former is based on a piecewise function to model the general conflicts with specific distribution while model the extreme conflicts with GPD. The threshold for distinguishing general conflicts and extreme conflicts is taken as a model parameter which needs to be estimated. Both two parts combine traffic conflicts from different sites and incorporates covariates and between-site unobserved heterogeneity into model parameters for addressing the non-stationarity of traffic conflict. The proposed model could determine the threshold in the EVT peak over threshold approach in an objective way.

The proposed BHHM framework was applied to model traffic conflicts from three signalized intersections in the city of Surrey, British Columbia. Traffic conflicts were characterized by the conflict indicator PET, and negative PET was used for model fitting. Five non-stationary BHHM models, including the Normal-GPD model, Cauchy-GPD model, Logistic-GPD model, Gamma-GPD model, and Lognormal-GPD model, were developed. For comparison, two Bayesian hierarchical non-stationary GPD models were developed using threshold calculated from the graphical diagnostics and quantile regression approaches, respectively. The results suggest that, *a*) the proposed non-stationary BHHM framework is able to determine the threshold in an objective and quantitative way; *b*) the threshold varies across cycles, which is captured by the non-stationary models; *c*) the Lognormal-GPD model outperform the other four hybrid structure models in terms of DIC; *d*) the crash estimates from the hybrid structure models are generally more accurate, with narrower 95% confidence intervals compared to the graphical diagnostic and quantile regression approaches; and *e*) the Lognormal-GPD model is the best model in terms of crashes estimation.

The proposed non-stationary BHHM framework offers significant advantages including, *a*) the model can objectively estimate the threshold parameter by utilizing all observations, eliminating the subjective process and providing a more reliable peak over threshold method for calculating crash risk; *b*) by incorporating covariates and between-site unobserved heterogeneity for non-stationary extremes (Coles et al., 2001, Mannering et al.,2016), the variation characteristic of threshold was captured by the proposed model, eliminating the limitation of using fixed threshold in crash estimation in dynamic traffic environments; *c*) as this technique enables adaptive threshold determination, it allows for real-time (e.g. signal cycle level) crash risk calculation that has essential application in ITS and CAV area.

This study can be extended in a few directions as follows. First, although the sample size was sufficient in this study, it is recommended to collect conflict data over a longer observation period to better reflect the true traffic conditions and cover the variability in traffic data. Further study should be conducted to compare the proposed BHHM approach with the data-driven methods using large samples of traffic conflicts (Hussain et al., 2022). Secondly, this study only used rear-end conflicts and crashes to validate the proposed modeling approach. Future research could extend the model by incorporating various types of conflicts (e.g. rear-end and side-swipe conflict) extracted from high-resolution vehicle trajectory data. Thirdly, previous studies have showed that incorporating multiple types of traffic conflict indicators in joint modeling can comprehensively assess the safety levels of investigated locations (Fu et al., 2020; Zheng and Sayed, 2020; Arun et al., 2021). Investigating the automatic determination of thresholds in bivariate and multivariate peak over threshold methods is worth exploring in future research. Additionally, the study locations are undersaturated signalized intersections, where queues are generally dissipated within a single signal cycle. However, as for the oversaturated signalized intersections, queues persist beyond a single cycle, leading to challenges such as infinite shockwave, presenting different traffic dynamics. As such, future studies should be conducted to investigate how the proposed approach can be extended to handle oversaturated intersections and explore how persistent queues could be incorporated into the modeling framework. Lastly, in addition to the peak over threshold method, the block maxima method is also popular for estimating crashes. The current method of determining the block size in the block maxima method also suffers from the issue of subjectivity. If the block size is too small, the fitting may be inaccurate, potentially leading to biased estimations. Conversely, if the block size is too large, only a few extreme values may be extracted, resulting in a high variance. Future research could focus on selecting an accurate and appropriate block size.

## Acknowledgements

This research is sponsored by the National Natural Science Foundation of China (Grant No. 51925801, 52232012, 52272343 & 52131203) and Postgraduate Research & Practice Innovation Program of Jiangsu Province (Grant No. KYCX24_0452).